\documentclass[12pt]{article}
\usepackage{latexsym}
\usepackage{amssymb}
\usepackage{amsmath}
\usepackage{graphicx}
\usepackage{hyperref}

\def\be{\begin{equation}}
\def\ee{\end{equation}}
\def\bear{\begin{eqnarray}}
\def\eear{\end{eqnarray}}
\def\nn{\nonumber}

\newcommand\bra[1]{{\langle {#1}|}}
\newcommand\ket[1]{{|{#1}\rangle}}

\def\a{\alpha}
\def\b{\beta}
\def\g{\gamma}
\def\d{\delta}

\def\r{\rho}

\def\z{\zeta}
\def\s{\sigma}
\def\th{\theta}

\def\Tr{{\rm Tr}}

\def\dd{\mbox{d}}

\def\O{\Omega}
\def\o{\omega}
\def\bra{\langle}
\def\ket{\rangle}
\def\a{\alpha}
\def\b{\beta}
\def\d{\delta}
\def\D{\Delta}

\def\g{\gamma}
\def\G{\Gamma}
\def\e{\epsilon}

\def\f{\phi}
\def\F{\Phi}
\def\vf{\varphi}

\def\l{\lambda}
\def\L{\Lambda}
\def\m{\mu}
\def\n{\nu}
\def\o{\omega}

\def\r{\rho}

\def\th{\theta}

\def\pa{\partial}

\newcommand{\ti}[1]{\tilde{#1}}

\newcommand{\tn}[1]{\mbox{\tiny #1}}
\renewcommand{\@}[1]{\sqrt{#1}}
\renewcommand{\le}[1]{\label{#1}\end{eqnarray}}
\newcommand{\bea}{\begin{eqnarray}}
\newcommand{\eea}{\end{eqnarray}}

\newcommand{\eq}[1]{(\ref{#1})}
\def\nn{\nonumber\\}

\def\ffract#1#2{\raise .35 em\hbox{$\scriptstyle#1$}\kern-.25em/
\kern-.2em\lower .22 em \hbox{$\scriptstyle#2$}}

\def\Ric{{\mbox{Ric}}}

\def\na{\nabla}
\def\half{{1\over2}\,}

\begin{document}
\pagestyle{empty}

\centerline{{\Large \bf The Invisibility of Diffeomorphisms}}
\vskip 1truecm

\begin{center}
{\large Sebastian De Haro}\\
\vskip .3truecm
{\it Trinity College, Cambridge CB2 1TQ, United Kingdom}\\
{\it Amsterdam University College, University of Amsterdam,
Science Park 113\\ 1090 GD Amsterdam, The Netherlands}

\vskip .1truecm
{\tt sd696@cam.ac.uk}
\vskip .5truecm 
\end{center}

\vskip 6.5truecm

\begin{center}

\textbf{\large \bf Abstract}
\end{center}

I examine the relationship between $(d+1)$-dimensional Poincar\'e metrics and $d$-dimensional conformal manifolds, from both mathematical and physical perspectives. The results have a bearing on several conceptual issues relating to asymptotic symmetries, in general relativity and in gauge-gravity duality, as follows:

(1:~Ambient Construction)~~I draw from the remarkable work by Fefferman and Graham (1985, 2012) on conformal geometry, in order to prove two propositions and a theorem that characterise the classes of diffeomorphisms that qualify as {\it gravity-invisible}. I define natural notions of gravity-invisibility (strong, weak, and simpliciter) which apply to the diffeomorphisms of Poincar\'e metrics in any dimension. 

(2:~Dualities)~~I apply the notions of invisibility
to gauge-gravity dualities: which, roughly, relate Poincar\'e metrics in $d+1$ dimensions to QFTs in $d$ dimensions. I contrast {\it QFT-visible} vs.~{\it QFT-invisible} diffeomorphisms: those gravity diffeomorphisms that can, respectively cannot, be seen from the QFT. 

The QFT-invisible diffeomorphisms are the ones which are relevant to the hole argument in Einstein spaces.  
The results on dualities are surprising, because the class of QFT-visible diffeomorphisms is {\it larger} than expected, and the class of QFT-invisible ones is {\it smaller} than expected, or usually believed, i.e.~larger than the PBH diffeomorphisms in Imbimbo et al.~(2000). I also give a general derivation of the asymptotic conformal Killing equation, which has not appeared in the literature before.

\newpage
\pagestyle{plain}

\tableofcontents

\newpage

\section{Introduction}\label{results}

The asymptotic symmetries of gravity have been a central foundational topic in general relativity since at least the work Arnowitt et al.~(1959, 2008), Sachs (1961, 1962), Bondi et al.~(1962), Penrose (1963, 1964), Newman et al.~(1966), Geroch (1972), Ashtekar et al.~(1978), and others. A central question is whether there are asymptotic diffeomorphisms that act on the physical degrees of freedom of the gravity theory, and how these diffeomorphisms are to be characterised. Only very recently has it for example been realized that, for Schwarzschild spacetimes, there are---in addition to the usual ADM mass, momentum, and angular momentum---an infinite number of conserved supertranslation and superrotation charges, which act non-trivially on the physical phase space (Hawking et al.~(2016)).

In this paper, I analyse the case of a negative cosmological constant. (For a discussion of the other cases: see the physical motivation, below.) I will use gauge-gravity duality to argue that there is a significant, non-empty, class of diffeomorphisms---which I will, broadly speaking, call `visible', in a sense that I will make precise---which act on the dual gauge theory, and which act on the physical degrees of freedom of the gravity theory. And there is a class of `invisible' diffeomorphisms which do not act on the asymptotic quantities. The latter class invites a comparison with Einstein's hole argument. 

I will develop techniques to characterise these two classes, and I will prove a theorem and two propositions about them.\\

{\it Diffeomorphisms and gauge-gravity duality.} Gauge-gravity dualities are surprising relationships between gravity theories, typically defined in $d+1$ dimensions, and quantum field theories (QFTs) in $d$ dimensions. The duality is usually construed as an `isomorphism' between all the physical quantities on either side. One important question for dualities is what part of the content of the theory is `physical', and thus mapped by the duality: and what part of content is `unphysical', specific to one of the two sides, hence not mapped by the duality---it will be {\it invisible} to duality. Gauge symmetries in QFT are of this kind: if the QFT has a gauge symmetry, its physical quantities are gauge invariant and are treated as such by the duality---the gauge symmetry is not seen on the dual side.

One naturally expects that the diffeomorphism invariance of the gravity theory is also of this kind: what is physical in a gravity theory should be independent of the coordinates chosen, and so one would naively not expect the duality to `see' the action of diffeomorphisms in the gravity theory. The QFT does {\it not} possess diffeomorphism invariance, and so the diffeomorphisms should be invisible to it. But there is one well-known class of diffeomorphisms that {\it is} visible through the duality map and which thereby can acquire a physical meaning (De Haro (2016a:~\S1.3.2)). Namely, the QFT is invariant under the coordinate transformations that leave its background geometry fixed. In the cases where the QFT has an UV fixed point (at which it is a conformal field theory, or CFT), the conformal group is known to arise, through the duality map, from a restricted class of diffeomorphisms of the gravity theory, which go under the name of PBH transformations (Brown and Henneaux (1986:~\S III-IV), Imbimbo et al.~(2000), De Haro et al.~(2001)). 

The difference between the two kinds of diffeomorphisms---those that are visible vs.~those that are invisible through the duality---is thus a crucial property of the duality map, and determines what is `physical', on both sides of the duality. The diffeomorphisms differ both in their physical properties and in the ways in which they can be regarded to be novel properties of the gravity theory. While I will leave the question of emergence of diffeomorphisms for the future:\footnote{For a discussion of emergence {\it of spacetime} in gauge-gravity dualities, see De Haro (2016:~\S3).} in this paper I will focus on the mathematical and physical contrast between visible and invisible diffeomorphisms.\\

{\it Physical motivation and generality of the results.} Let me describe in more detail the two main physical motivations for this work: namely, from general relativity, and from quantum gravity.

As for classical general relativity: there is, of course, a large and venerable literature on boundary conditions, and diffeomorphisms which preserve them, in general relativity: especially in the {\it asymptotically flat} case. Arnowitt et al.~(1959, 2008) developed the definition of energy using the ADM formalism, in which spacetime is foliated into a family of spacelike surfaces, and they parametrised the four-dimensional metric in terms of a three-dimensional metric on the surface and four functions, the lapse function and the shift vector. Sachs (1961, 1962) and Bondi et al.~(1962) studied in detail the question of asymptotic symmetries at null infinity in asymptotically flat spacetimes, a problem that is highly relevant to e.g.~gravitational waves. The asymptotic symmetry group discovered now goes under the name of the BMS group. This led to other important results, such as Penrose's (1963, 1964) treatment of conformal infinity, which also holds in the presence of a non-zero cosmological constant. The asymptotically flat case was further developed in works such as Newman et al.~(1966), Geroch (1972), Ashtekar et al.~(1978), and others. 

The case of a {\it negative cosmological constant} has been treated, with a variety of motivations, in the works cited in the preamble of this Introduction. Other important work is e.g.~Ishibashi et al.~(2004), which focuses on AdS's lack of global hyperbolicity. 

The case of a {\it positive cosmological constant} is the poorest understood. Relevant works are e.g.~Anninos et al.~(2011) and Ashtekar et al.~(2015), and references therein.

While the cosmological constant in our universe is of course not negative\footnote{I thank an anonymous referee for bringing up the question of the relevance of this work for actual cosmology.} (nor is it zero!), there are several motivations, from classical general relativity, for taking up the case of a negative cosmological constant once again: in addition to the ones already mentioned earlier. 

First of all, as in Ishibashi et al.~(2004), the case of negative cosmological constant is non-globally hyperbolic (since pure AdS is ``like a box''), and so understanding in detail how to define boundary conditions, and how boundary conditions and diffeomorphisms mesh, is quite relevat for the treatment of solutions more generally in open regions of the universe, where observers within any finite region have no access to infinity within a finite time. And so, it is of conceptual and practical importance to understand general relativity for open systems (the Schwarzschild black hole being a related example). 

Second, the techniques which I develop in this paper can be generalised, by an analytic continuation $\ell_{\tn{AdS}}\mapsto i\,\ell_{\tn{dS}}$, to the cosmologically relevant case of a {\it positive} cosmological constant: as I discuss towards the end of Section \ref{discussionandc} (for details on how this map acts, see De Haro et al.~(2016a:~\S8)). The analytic continuation maps the timelike boundary to a spacelike boundary. In fact, one expects not only the mathematical techniques, but also some of the conceptual lessons, to carry over to that case: such as the bulk/hole argument of De Haro et al.~(2016: Section 6), and the notion of gravity-invisible diffeomorphisms. 

But there is of course also, in addition to these classical considerations, a quantum gravity motivation: understanding the classical structure of gauge-gravity duality is an important step towards understanding the duality at the quantum level. For asymptotic symmetry structures are of course important for the quantisation of gravity. Since candidate quantum gravity theories do not abound, developing AdS/CFT is a worthwhile exercise. And as stressed in De Haro (2016a): the content that is invariant across the duality (the `common core') is what should be regarded as physically signficant for this particular theory of quantum gravity. This gives us an additional argument to the effect that the diffeomorphisms which are visible to the QFT also act on general relativity's asymptotic degrees of freedom. 

The question, of which class of diffeomorphisms are physical and which are unphysical, is an important question for dualities in general---as it is for gauge theories. It also bears on the definition of observables, background-independence, and emergence. Thus AdS/CFT is a good case study which has already provided insights into possibilities for defining a gauge-gravity duality for spaces with a positive cosmological constant (see e.g.~Maldacena (2003), Strominger (2001), De Haro et al.~(2016a:~\S8)). 

\subsection{Conformal geometry and summary of the results}

I will draw on the so-called {\it ambient construction} in conformal geometry---a remarkable piece of mathematics by Fefferman and Graham (1985, 2012)---in order to prove two propositions and a theorem which apply to general relativity and gauge-gravity dualities. The mathematical results concern the conditions under which a diffeomorphism, in a gravity theory with a gauge dual, is `invisible' to the gauge theory.\footnote{The notion of `invisibility' for dualities in general was introduced in De Haro, Teh, and Butterfield (2016:~\S5.1). It is concretely inspired by the work of Horowitz and Polchinski (2006). See the last paragraph of this Section, and especially De Haro et al.~(2016:~\S2,5.1-5.2),  for a discussion.} I will provide four notions of invisibility, three concerning the gravity theory and one concerning the gauge theory. The notions of gravity-invisibility amount to a diffeomorphism being invisible if it fixes certain mathematical structures in the gravity theory:\\
\indent(i) the form of the metric: i.e.~a class of Poincar\'e metrics,\\
\indent(ii) the conformal manifold at the boundary: in terms of its {\it points}, or \\
\indent(iii) the representative of the conformal class of metrics with which the boundary manifold is equipped.

As we will see in Section \ref{visin}, fixing (ii) does not imply fixing (iii): for the class of diffeomorphisms fixing (ii) include non-trivial conformal transformations at the boundary, which transform the representative of the conformal class non-trivially, hence do not fix (iii).

I will define notions of invisibility that apply to the two theories involved in a gauge-gravity duality in a moment. 

Let a {\it $T$-invisible diffeomorphism} be a diffeomorphism that is invisible to theory $T$, in the sense of its preserving appropriate structures of theory $T$. Let us now proceed to specify these structures in more detail.

For the gravity theory, there are three related notions of {\it gravity-invisibility}, depending on which of the structures (i)-(iii) above are preserved, as follows: \\
\indent(a) {\it strongly gravity-invisible} diffeomorphisms: which fix all of (i)-(iii); \\
\indent(b) {\it weakly gravity-invisible} diffeomorphisms: which fix (i) \& (ii) or (i) \& (iii) but not necessarily all three; \\
\indent(c) (simpliciter) {\it gravity-invisible} diffeomorphisms: which fix (ii) \& (iii) but not necessarily (i). 

The notion of QFT-invisible diffeomorphisms, on the other hand, concerns the QFT: they are those gravity diffeomorphisms which cannot be seen (in a sense yet to be made precise) through the duality, hence are invisible to the QFT. 

Thus, my definition of `invisibility of diffeomorphisms' is relative to a theory (the gravity theory or the QFT): more precisely, relative to certain structures preserved within that theory. Thus the gravity-invisible diffeomorphisms are a priori independent of the duality, and express only a property of the gravity theory. The QFT-invisible diffeomorphisms will be the ones that should be seen as a property of the duality, viz.~they are diffeomorphisms of the gravity theory which are invisible to the QFT (and they will be defined in terms of gravity-invisible diffeomorphisms).

The main mathematical results of this paper can then be summarised in the following three statements regarding infinitesimal diffeomorphisms (keeping the same numbering (a)-(c), since each result refers to its corresponding class above):\\
\indent(a: Theorem 3, \S\ref{inficase}) There exist no non-trivial {\it strongly gravity-invisible diffeomorphisms}, i.e.~imposing that the diffeomorphism is strongly gravity-invisible also implies that it is equal to the identity. \\
\indent(b: Propositions 1-2, \S\ref{inficase}) The {\it weakly gravity-invisible diffeomorphisms} reduce to conformal transformations at the boundary of the manifold.\\
\indent(c: \S\ref{invisible}) There exist non-trivial {\it gravity-invisible diffeomorphisms}. \\
\\
\indent These mathematical results have a number of surprising physical and philosophical consequences:\\

(1) There is a version of Einstein's hole argument for (generalised) anti-de Sitter (AdS) space: what we may call the `bulk argument', introduced in De Haro, Teh, and Butterfield (2016:~Section 6). The result (c) in the current paper implies that there is indeed a {\it non-empty} class of diffeomorphisms for which the bulk argument holds. And result (c) also characterises this class: as being smaller than one might expect.\\

\indent(2) The weakly gravity-invisible diffeomorphisms (b) give rise to the conformal symmetry of the gauge theory, with the implication that not all diffeomorphic structure in the gravity theory is invisible to the quantum field theory (QFT). This substantiates the claim in De Haro, Teh, and Butterfield (2016:~\S5.1) that not all `gauge' structure (in the philosopher's sense) is invisible to the duality. Although the connection between the diffeomorphisms in the gravity theory and conformal invariance is familiar from the gauge-gravity literature, the class of diffeomorphisms which give rise to conformal transformations is in this paper found to be larger than the standard one in Imbimbo et al.~(2000) and Skenderis (2001): see the discussion following Eqs.~\eq{dg0} and \eq{conftr}. \\

\indent(3) The distinction between visible and invisible diffeomorphisms, worked out in mathematical detail here, underlies the discussion of background-independence in De Haro (2016:~\S\S2.3.2-2.3.4): and, in particular, it characterises two classes of diffeomorphisms to which a different analysis of background-independence applied, in \S2.3.3 of De Haro (2016). In that paper, these two cases were distinguished from each other and from yet antoher class, of `large' diffeomorphisms: which do not preserve any of the pairwise structures defined here, and which I will not consider in this paper. The distinction of QFT-visibility vs.~QFT-invisibility also  provides the basis of the discussion, in \S2.3.3 of De Haro (2016), of the purported covariance of states and quantities. The violation of covariance for even boundary dimensions is given in Eq.~\eq{anomalous}. \\

\indent(4) Having a precise characterisation of the notions of visibility and invisibility of diffeomorphisms, it now becomes possible to meaningfully discuss whether, and how, diffeomorphisms emerge on the gravity side. One point that readily follows from (a)-(c) is that, despite the claims in the literature, there is no `emergence of diffeomorphisms' {\it tout court}: for the visible and the invisible diffeomorphisms do not arise in anything like the same sense. I shall leave this question for the future.

My results provide a completely general derivation of the condition for a gravity diffeomorphism to give rise to a conformal transformation on the boundary, which, though perhaps known to the experts in the geometry of gauge-gravity dualities,\footnote{This was confirmed in: K.~Skenderis, private communication.} has not appeared in print except in very special cases. So, the results here fill a gap in the literature: indeed, to my knowledge, the derivation of the condition for the diffeomorphisms to be conformal transformations, i.e.~the gravity derivation of the QFT's conformal Killing equation from the requirement of weak invisibility (Eq.~\eq{dg0} for the linear case, Eq.~\eq{ijirrr2} for the non-linear case) has not appeared in the literature except for pure AdS (Gubser, Klebanov, and Polyakov (1998:~Eq.~(18))) and low-dimensional cases (Brown and Henneaux (1986:~\S III-IV)).

\subsection{Plan of the paper}

In  Section \ref{visin}, I introduce and develop the methods from conformal geometry that are needed to be able to define visibility and invisibility with the precision required for our purposes. I then prove the results (a)-(c), which provide the mathematical basis for: (1) and (2), which were discussed in De Haro, Teh, and Butterfield (2016); as well as (3), which was discussed in De Haro (2016). Three Appendices contain technical and illustrative examples of the relevant physics, and of how QFT-invisibility shows in these examples.

The notion of invisibility is motivated by a discussion by Horowitz and Polchinski (2006:~p.~12): `the gauge theory variables... are trivially invariant under the {\it bulk diffeomorphisms, which are entirely invisible in the gauge theory}' (my emphasis). It follows from the analysis in the current paper that not {\it all} gravity diffeomorphisms are in fact invisible to the gauge theory. As we saw in (c) above, there is a large class (larger than normally realised\footnote{`Normal' here refers to the standard references, in the context of gauge-gravity duality, on the so-called PBH transformations: Imbimbo et al.~(2000), De Haro et al.~(2001), Skenderis (2001). See the discussion following Eqs.~\eq{dg0} and \eq{conftr}.}) of diffeomorphisms of the gravity theory which are {\it not} invisible to the gauge theory: those that do not restrict to the identity map on the boundary, under which the gauge theory is not invariant but {\it co}variant at best (in the case of odd $d$), and non-invariant (because of an anomaly when $d$ is even) at worst. Section \ref{ginvis} will specify the class of QFT-invisible diffeomorphisms: the specification of the class turns out to be subtle, and the class turns out to be smaller than often expected. In Section \ref{discussionandc}, I discuss and summarise the results.

Though I take the discussion by Horowitz and Polchinski as my {\it motivation} for considering invisibility, my definition of the notion differs from theirs, in that, as mentioned in the preamble of this Section, it is relative to a specific theory: and so, I allow for diffeomorphisms that are invisible not only to the gauge theory, but also for diffeomorphisms that are invisible to the gravity theory (in the sense that they preserve the structures (i), (ii) or (iii)).

\section{Visible vs.~Invisible Diffeomorphisms}\label{visin}

In this Section, I prove the main mathematical results of the paper, (a)-(c) in Section \ref{results}, concerning three kinds of gravity-invisible diffeomorphisms. In \S\ref{PNF}, I will collect the definitions and theorem, from Fefferman and Graham (1985, 2012), that will be used in the rest of the section. In \S\ref{invisibility}, I will define the relevant notions of invisibility and derive two propositions and our main theorem about them: (a) that the class of non-trivial strongly-invisible diffeomorphisms is empty, as well as (b) weakly gravity-invisible diffeomorphisms reduce to boundary conformal transformations. In \S\ref{invisible}, I will prove that (c) the class of non-trivial gravity-invisible diffeomorphisms is non-empty and I will give bounds on the asymptotic behaviour that ensure that such diffeomorphisms in fact exist. I will use these results in Section \ref{ginvis} to define the notion of QFT-invisible diffeomorphisms, and I will explain how it relates to the gravity-invisible diffeomorphisms. 

Throughout, we will be considering solutions of Einstein's equation in $d+1$ dimensions in vacuum\footnote{Appendix \ref{matter} discusses how to couple gravity to matter fields.} with a negative cosmological constant $\L=-{d(d-1)\over2\ell^2}$, where $\ell$ is called the {\it curvature radius}:
\bea\label{Einstein}
\Ric[\hat g]+{d\over\ell^2}\,\hat g=0~,
\eea
and $\hat g$ is the $(d+1)$-dimensional metric (as opposed to $g$, which will denote a $d$-dimensional metric: to be defined below) of any signature.


\subsection{Poincar\'e metrics and normal forms}\label{PNF}

Our aim in this subsection is to introduce the geometrical notions that will allow us to articulate, in \S\ref{invisibility}, three related notions of invisibility of a diffeomorphism. To this end, I will first, in \S\ref{confman}, introduce conformal manifolds. Then I will define the notion of conformal compactness: manifolds whose metric, roughly speaking, has a double pole at the boundary, but is otherwise smooth and nondegenerate at the boundary, which is itself a conformal manifold. Then I will require the metric on this conformally compact manifold to be of  Poincar\'e type, and introduce some results about the normal form of this metric. In \S\ref{diffeo}, I will discuss diffeomorphisms, both active and passive: which will allow us to discuss their invisibility in \S\ref{invisibility}.

\subsubsection{Conformal manifolds and Poincar\'e metrics}\label{confman}

{\bf Definitions.}\footnote{The definitions and conventions in this subsection mostly follow Fefferman and Graham (2012).} A {\bf conformal structure} on a differentiable manifold $M$ is an equivalence class of (pseudo)-Riemannian metrics, in which two metrics are equivalent if one is a positive smooth multiple of the other. We will denote a {\bf conformal class}, i.e.~such a conformal structure, by $[g]$. Thus, $[g]$ consists of all metrics on $M$ of the form $\Omega^2\, g$, where $\Omega$ is any smooth, real-valued function on $M$. $g$ is a smooth metric, called a {\bf representative} of the conformal class $[g]$.

Throughout this paper, $M$ will be a smooth manifold of dimension $d\geq2$, equipped with a conformal structure $[g]$. The representative $g$ of the class will be a smooth pseudo-Riemannian metric of signature $(p,q)$ on $M$, with $p+q=d$. 
A {\bf conformal manifold}, then, is a pair $(M,[g])$ of a smooth manifold of dimension $d\geq2$, equipped with a conformal structure, which is a choice of a conformal class of metrics of signature $(p,q)$. 

Let $\hat M$ be a manifold with boundary $M$, $\pa\hat M=M$. Pick a defining function for this boundary: a function $r\in C^\infty(\hat M)$ which satisfies: (i) $r>0$ in the interior $\hat M_{\tn{int}}=\hat M-M$, (ii) $r=0$ on $M$, and: (iii) $\dd r\not=0$ on $M$. 

We will be concerned with the behaviour near the boundary $M$ of $\hat M$. Locally near $r=0$, $\hat M$ has the form of a product manifold. Thus we will consider an open neighbourhood of $M\times\{0\}\subset M\times\mathbb{R}_{\geq0}$, where the defining function $r\in\mathbb{R}_{\geq0}$ denotes the second factor.\\
\\
{\bf Definition.} A smooth metric $\hat g$ on the interior of $\hat M$, $\hat M_{\tn{int}}$, of signature $(p+1,q)$ is {\bf conformally compact}, if: (i) $r^2\hat g$ extends smoothly to $\hat M$, and: (ii) $r^2\hat g|_M$ is nondegenerate (i.e.~of signature $(p+1,q)$ also on $M$).
A conformally compact metric $\hat g$ is said to have {\bf conformal infinity} $(M,[g])$ if $r^2\hat g|_{TM}\in[g]$.\\
\\
{\bf Definition} (Fefferman and Graham (2012:~\S4.1)). A {\bf Poincar\'{e} metric} for $(M,[g])$ is a conformally compact metric $\hat g$ of signature $(p+1,q)$ on $\hat M_{\tn{int}}$, where $M_{\tn{int}}$ is an open neighbourhood of $M\times\{0\}\subset M\times\mathbb{R}_{\geq0}$, such that:\\
\indent(1) $\hat g$ has conformal infinity $(M,[g])$. \\
\indent(2) If $d$ is odd or $d=2$, then $\Ric[\hat g]+{d\over\ell^2}\,\hat g$ vanishes to infinite order along $M$.\\
\indent If $d\geq4$ is even, then $\Ric[\hat g]+{d\over\ell^2}\,\hat g={\cal O}(r^{d-2})$, i.e.~$\Ric[\hat g]+{d\over\ell}\,\hat g$ `vanishes up to terms of order' $r^{d-2}$.\\
The same results apply if one considers metrics $\check{g}$ on $\hat M_{\tn{int}}$ of signature $(p,q+1)$ such that $\Ric[\check g]-{d\over\ell^2}\,\check g$ vanishes to the stated order. \\
\\
{\bf Definition} (based on Fefferman and Graham (2012:~\S4.2)). A Poincar\'{e} metric $\hat g$ for $(M,[g])$ is said to be in {\bf normal form} relative to $g$ if:
\bea\label{NF}
\hat g={\ell^2\over r^2}\left(\dd r^2+g_r\right),
\eea
where $g_r$ is a 1-parameter family of metrics on $M$ of signature $(p,q)$, such that $g_0=g$.

There is an alternative form of a Poincar\'e metric in normal form, with formal asymptotics that is entirely equivalent. It is suggested by Fefferman and Graham's (1985) ambient space construction that originally motivated their work. There is a diffeomorphism $\chi_\ell:M\times\mathbb{R}_{\geq0}\rightarrow M\times\mathbb{R}_{\geq0}$, $\chi_\ell(x,r)=\left(x,\sqrt{\ell\r}\right)$ bringing the above metric to the following form:
\bea\label{FG}
\hat g={\ell^2\over4\r^2}\,\dd\r^2+{\ell\over\r}\,g(x,\r)~,
\eea
where $g(x,\r)=g_{\sqrt{\ell\r}}(x)$ is a 1-parameter family of metrics on $M$ satisfying $g(x,0)=g(x)=g_{ij}(x)\,\dd x^i\,\dd x^j\in [g]$, for a coordinate system $(x^1,\ldots,x^d)$ on $M$. \\
\\
{\bf Theorem.} (Fefferman and Graham (2012:~\S4.5)). Let $M$ and $g$ be given as above. Then there exists an even (i.e.~it is an even function of $r$) Poincar\'e metric $\hat g$ for $(M,[g])$ which is in normal form relative to $g$.

\subsubsection{Diffeomorphisms}\label{diffeo}

Let $p$ be a point in a neighbourhood $\,{\cal U}_1$ of $\hat M$. Let $\varphi$ be a coordinate function on $\,{\cal U}_1$, i.e.~there is a chart $(\,{\cal U}_1,\varphi)$, such that $\varphi:~\,{\cal U}_1\rightarrow\mathbb{R}^{d+1}$, viz.~it assigns $p\mapsto \varphi(p)$. Call the point that $\varphi$ maps to, $X:=\varphi(p)\in\mathbb{R}^{d+1}$. Let $\,{\cal U}_2$ be another neighbourhood of $\hat M$ with coordinate chart $(\,{\cal U}_2,\psi)$, such that $\psi:~\,{\cal U}_2\rightarrow\mathbb{R}^{d+1}$, viz.~an assignment $q\mapsto \psi(q)$. Call the point that $\psi$ maps to, $\ti X:=\psi(q)\in\mathbb{R}^{d+1}$. 

A {\bf diffeomorphism} $\f:~\,{\cal U}_1\rightarrow\,{\cal U}_2$ is a homeomorphism that assigns to $p$ another point $q=\f(p)$, $\f:p\mapsto\f(p)$, such that the map $\F:=\psi\circ\f\circ\varphi^{-1}:~\mathbb{R}^{d+1}\rightarrow\mathbb{R}^{d+1}$ between the respective coordinates, i.e.~$(\F\circ\varphi)(p)=(\psi\circ \phi)(p)$, is invertible, and both $\F$ and $\F^{-1}=\varphi\circ\f^{-1}\circ\psi^{-1}$ are $C^\infty$. We can also write this condition in terms of invertibility and differentiability of the function $\ti X=\F(X)$ on $\mathbb{R}^{d+1}$ and its inverse $X=\F^{-1}(\ti X)$. 

When $\,{\cal U}_1=\,{\cal U}_2$, so that $\f:~\,{\cal U}\rightarrow\,{\cal U}$, we can take $\psi=\vf$ and $\F=\psi\circ\f\circ\psi^{-1}$. Then $X$ and $\ti X$ correspond to different points in $\,{\cal U}$, in the same coordinate chart. Such a diffeomorphism is called {\bf active}. In this paper we will construe all diffeomorphisms as active.

One can also consider {\bf passive} diffeomorphisms, which are mere reparametrizations of the coordinates: one considers a single point $p$ and two overlapping coordinate charts $\left(\,{\cal U}_1,\vf\right)$, $\left(\,{\cal U}_2,\psi\right)$ such that $p\in \,{\cal U}_1\cap\,{\cal U}_2$. The map $\F:\mathbb{R}^{d+1}\rightarrow\mathbb{R}^{d+1}$, $\vf(p)\mapsto \F(\vf(p))=\psi(p)$, in other words $\F(X)=\ti X$, is then taken to be differentiable. The formula is the same, but the meaning of the diffeomorphism is different: since $X$ and $\ti X$ now correspond to the {\it same} point $p\in\,{\cal U}_1\cap\,{\cal U}_2$, but expressed in different coordinate charts. \\
\\
{\bf Proposition (Diffeo)} (Fefferman and Graham (2012:~\S4.3)). Let $\hat g$ be a Poincar\'e metric on $\hat M_{\tn{int}}$ for $(M,[g])$. Then there exists an open neighbourhood ${\cal U}$ of $M\times\{0\}\subset M\times\mathbb{R}_{\geq0}$ on which there is a unique diffeomorphism $\f:\,{\cal U}\rightarrow \hat M$ such that $\phi|_M$ is the identity map, and $\f^*\hat g$ is in normal form relative to $g$ on $\,{\cal U}$.\

So, when we work with Poincar\'e metrics, we only need to consider those that are in normal form.

\subsection{Strongly gravity-invisible diffeomorphisms are the identity}\label{invisibility}

In this subsection, I will introduce three related notions of gravity-invisibility, and prove my main results about them, viz.~(a)-(c) in Section \ref{results}: \\
\indent(a) non-trivial strongly gravity-invisible diffeomorphisms do not exist;\\
\indent(b) weakly gravity-invisible diffeomorphisms reduce to boundary conformal transformations; \\
\indent(c:~in \S\ref{invisible}) there exist non-trivial gravity-invisible diffeomorphisms.

Consider a Poincar\'e metric $\hat g$ for $(M,[g])$. By (Diffeo), we take this metric to be in normal form relative to $g$ in an open neighbourhood $\,{\cal U}$ of $M\times\{0\}\subset M\times\mathbb{R}_{\geq0}$.\footnote{There is of course no claim here that $\f$ in (Diffeo) is invisible. Grumiller et al.~(2016:~Eq.~(3.6)) report a three-dimensional metric that is claimed to be physically inequivalent to the corresponding metric in normal form.}

Now consider a diffeomorphism $\f$ of the manifold, and the pullback $\f^*\hat g$ of the metric that it gives rise to. 
Let $\f:~\,{\cal U}\rightarrow\,{\cal U}$ be a diffeomorphism, defined as in \S\ref{diffeo}. We will be interested in the class of diffeomorphisms that preserve the normal form of the metric. We will also impose various conditions on the asymptotic form of the diffeomorphism. This will be encapsulated in the idea of a diffeomorphism being {\it invisible} (in one or another of three related senses); and our first aim, roughly speaking, will be to prove that only the identity diffeomorphism is invisible.
As mentioned, we will consider active diffeomorphisms, though similar considerations apply to the passive ones. Thus we set $\psi=\vf$ in the definition of an active diffeomorphism, in \S\ref{diffeo}. Let us start with some definitions.\\
\\
{\bf Definition.} Let $\hat g$ be a Poincar\'e metric for $(M,[g])$ in normal form. A diffeomorphism $\f:\,{\cal U}\rightarrow\,{\cal U}$, where $\,{\cal U}$ is an open neighbourhood of $M\times\{0\}\subset M\times\mathbb{R}_{\geq0}$, is said to be {\bf invisible relative to} $(\hat g,M,g)$ (or {\bf strongly gravity-invisible}) if it satisfies the following three conditions:\\
\\
(i)~\, (Invisible relative to $\hat g$)~~~:~~$\f^*\hat g$ is in normal form relative to $g$.\\
(ii) \,\,(Invisible relative to $M$)~:~~$\f|_{M\times\{0\}}=\mbox{id}_{M\times\{0\}}$. This means that $\F(x,0)=(x,0)$.\\
(iii) (Invisible relative to $g$)~\,\, :~~$(\f^*g)(p)=g(p)$, i.e.~$\f$ is an isometry of $M$. \\
\indent $\, $ In (iii), $p\in M$ and $(\f^*g)(p)$ is induced from $(\f^*g_r)(p)=g_{\ti r}(\f(p))$ at $r=0$ ($g=g_0$ in \eq{NF}), where $\ti r:=\Phi^{d+1}(x,r)$, the last component of $\Phi(x,r)\in\mathbb{R}^{d+1}$, which in what follows we shall denote $\Phi^r(x,r)$. Also, notice that (iii) is not trivially implied by (ii): for (ii) allows a non-trivial transformation of $r$, which we will parametrise as $\xi(x)$, and which is non-zero at the boundary and does transform $g$; whereas (iii) is the requirement that $g$ does not transform.\\
\\
If, under the above stated conditions, $\phi$ is invisible relative to $(M,g)$, in the sense that (ii) and (iii) hold but not necessarily (i), then $\phi$ is said to be {\bf gravity-invisible}. 

We will also consider diffeomorphisms that are invisible relative to $(\hat g,g)$ (i.e.~(i) and (iii) hold but not (ii) necessarily) or invisible relative to $(\hat g, M)$ (i.e.~(i) and (ii) hold but not (iii) necessarily): such $\phi$'s shall be collectively called {\bf weakly gravity-invisible} (and it will not be important for us to distinguish between the latter two conditions).

{\it Strongly gravity-invisible} vs.~{\it gravity-invisible} will be the crucial contrast for our discussion in \S\ref{invisibleq}-\ref{visibled}. 
Also, in this Section we will prove that a strongly gravity-invisible diffeomorphism must be the identity. The proof does not use (Diffeo) but it will be based on two propositions that (a) are interesting for their own sake, and (b) will give us insight into the the notion of invisibility.\\
\\
{\bf Definition.} A diffeomorphism $\vf_M$ on a manifold $M$ is called a {\bf conformal transformation} if its effect on the metric is to rescale it by some smooth, strictly positive function $\o:M\rightarrow\mathbb{R}_{>0}$, such that $(\varphi_M^*\,g)(p)=\o^{-2}(p)\,g(p)$.\\
\\
{\bf Definition.} Let $\hat g$ be a Poincar\'e metric for $(M,[g])$ in normal form. A diffeomorphism $\f:\,{\cal U}\rightarrow\,{\cal U}$, where $\,{\cal U}$ is an open neighbourhood of $M\times\{0\}$, is said to be a {\bf boundary-conformal diffeomorphism} (or simply, to be boundary-conformal) if $\f$ induces a conformal transformation on $g$, i.e.~there is a smooth, strictly positive function $\O:\hat M\rightarrow\mathbb{R}_{>0}$ such that:
\bea
\f^*\hat g(p)|_{p\in M}=\O^{-2}(p)\,\hat g(p)~.
\eea
{\bf Definition.} We will say that a diffeomorphism $\f$ on $\hat M$ {\bf reduces to a boundary diffeomorphism} $\varphi_M$ on $M$ if $\phi|_{M\times\{0\}}=\varphi_M\times\mbox{id}_{\{0\}}$.

Written in a coordinate patch, a diffeomorphism that reduces to a boundary diffeomorphism is one that satisfies: $\F(x,0)=(\ti x,0)$, where $\ti x=\psi(\vf_M(p))$ for $p\in M\subset M\times\{0\}$, and $x=\psi(p)$. This can be written as $\ti x=\Psi(x)$ where $\Psi:=\psi_M\circ\vf_M\circ\psi_M^{-1}:~\mathbb{R}^d\rightarrow\mathbb{R}^d$ and $\psi_M:=\psi|_M:~M\rightarrow\mathbb{R}^d\subset \mathbb{R}^d\times\{0\}$. 

Notice that a diffeomorphism that reduces to the identity on $M$ is invisible relative to $M$, i.e.~it trivially satisfies condition (ii) above.\\
\\
Let us also make a choice of coordinates on $\mathbb{R}^{d+1}$ in terms of which we will write the metric in the normal form \eq{NF}. Define $(x,r):=X=\psi(p)$ and $(\ti x,\ti r):=\ti X=\psi(\f(p))$. $\F$ is an invertible map. The diffeomorphism $X=\F^{-1}(\ti X)$ is then written:
\bea\label{ctr}
x^i&=&\left(\F^{-1}\right)^i(\ti x,\ti r)\nn
r&=&\left(\F^{-1}\right)^r(\ti x,\ti r)~,
\eea
where the superscript $r$ denotes the $(d+1)$-th component.

In the rest of this section we will be considering diffeomorphisms that are either invisible relative to $M$, or reduce to a boundary diffeomorphism $\varphi_M$. In both cases, the diffeomorphism acts as the identity on the second factor of $M\times\{0\}$. This means that, in both cases, $r=0$ and $\ti r=0$ each still parametrise the boundary. We will say that such a diffeomorphism {\bf fixes the location of the boundary}.\\
\\
{\bf Comment on the identity map.} Our condition (ii) of invisibility relative to $M$ is  $\phi_{M\times\{0\}}=\mbox{id}_{M\times\{0\}}$, implying that $\ti x=x$ and $\ti r=0$. Thus these diffeomorphisms fix the points of $M$ {\it at} $r=0$. This is a weaker condition than requiring that the diffeomorphism should go to the identity {\it in a neighbourhood} $U:=M\times[0,\e)$, for $\e>0$, of $M\times\{0\}$, i.e.~$\phi|_{U}=\mbox{id}_U$. The latter condition is stronger than (ii), and the former allows for diffeomorphisms which act nontrivially along the $r$-direction, $\ti r=\l(x)\, r$, while fixing $r=0$. Such diffeomorphisms generate conformal transformations at the boundary, as we will see in Propositions 1 and 2, thus they do not fix $g(p)$: and hence they do not imply (iii). 

\subsubsection{Infinitesimal case}\label{inficase}

In this section we will consider infinitesimal diffeomorphisms, as follows:\\
\\
(Infinitesimal) We only consider maps close to the identity map  in $\,{\cal U}$: $\f=\mbox{id}_{\,{\cal U}}+\d\f+\ldots$ Written out for $\F$, this  means that $\F={\mbox{id}}_{\mathbb{R}^{d+1}}+\psi\circ\d\f\circ\psi^{-1}+\ldots=:\mbox{id}_{\mathbb{R}^{d+1}}+\d\F+\ldots$ in $\vf(\,{\cal U})$. In the coordinates \eq{ctr}, we will write:
\bea\label{xxtilde}
x^i=\left(\F^{-1}\right)^i(\ti x,\ti r)&=&\ti x^i+\xi^i(\ti x,\ti r)\nn
r=\left(\F^{-1}\right)^r(\ti x,\ti r)&=&\ti r-\ti r\,\xi(\ti x,\ti r)~,
\eea
where $\xi^i$ and $\xi$ will be taken to be infinitesimal, and we will linearise all expressions in terms of them. 

If an infinitesimal diffeomorphism is to fix the boundary, then we immediately find that $\xi(\ti x,0)$ must be regular near $\ti r=0$ on $\psi(\,{\cal U})$, i.e.~$\xi(\ti x,\ti r)=\ti r^\a\,\xi(\ti x)+{\cal O}(\ti r^{\a+1})$ for some $\a\geq0$. The notation ${\cal O}(\ti r^{\a+1})$ means `up to terms of order $\ti r^{\a+1}$ and higher'. We will take the lowest value of $\a$ possible, viz.~$\a=0$, so that to account for higher values of $\a$ one simply sets $\xi(\ti x)=0$. Thus $r$ can be written as:
\bea
r&=&\ti r~\o(\ti x)+{\cal O}(\ti r^2)=
\ti r\left(1-\xi(\ti x)\right)+{\cal O}(\ti r^2)~,\label{r}
\eea
for $\o(\ti x)$ and $\xi(\ti x)$ both smooth functions. 

Let us now consider diffeomorphisms that are invisible relative to $\hat g$, i.e.~$\f^*\hat g$ is in normal form relative to a metric $g$ on the boundary manifold $M$. So, from \eq{NF}, for a point $q=\f(p)\in \,{\cal U}$, the following must hold:
\bea
\left(\f^*\hat g\right)_{ij}(p)&=&{\ell^2\over\ti r^2}~\tilde g_{ij}(q)\label{ij}\\
\left(\f^*\hat g\right)_{ir}(p)&=&0\label{ir}\\
\left(\f^*\hat g\right)_{rr}(p)&=&{\ell^2\over\ti r^2}~.\label{rr}
\eea
We will work out these three equations linearising in the diffeomorphisms $\d\F$, as in (Infinitesimal).

Equation \eq{rr} reduces to: $\left(\pa r\over\pa\ti r\right)^2{1\over r^2}={1\over\ti r^2}$. This can be integrated over the entire $\psi(\,{\cal U})$
\bea\label{r'}
r=\ti r~\o(\ti x)+{\cal O}(\xi_i^2)~.
\eea
So the lowest-order expression that we obtained in \eq{r} by assuming that $\xi(\ti x,\ti r)$ was regular at $r=0$, is actually valid on the entire domain $\varphi(\,{\cal U})$. 

Next we write out \eq{ir}. For this purpose, we use the just-obtained \eq{r'}. We get the following result:
\bea\label{ir'}
\pa_{\ti r}\xi^i(\ti x,\ti r)=\ti r\,g^{ij}(\ti x,\ti r)\,\pa_j\xi(\ti x)+{\cal O}(\xi^2,\xi_i^2)~.
\eea
The reason for the dependence of $g^{ij}$ on $(\ti x,\ti r)$ rather than $(x,r)$ is that the expression is already linear in $\xi, \xi^i$, so $(x,r)$ can be replaced with $(\ti x,\ti r)$ in the entire equation.

Finally we work out \eq{ij}, again for infinitesimal diffeomorphisms:
\bea\label{gtilde}
\ti g_{ij}(\ti x,\ti r)=\left(1+\xi(\ti x)\left(2-\ti r\,\pa_{\ti r}\right)\right)g_{ij}(\ti x,\ti r)+\nabla_i(g)~\xi_j(\ti x,\ti r)+\nabla_j(g)~\xi_i(\ti x,\ti r)+{\cal O}(\xi^2,\xi_i^2)~.
\eea
It will be useful for later use to write this as:
\bea\label{dg}
\d_{\f^{-1}} \,g_{ij}(x,r)&:=&(\f^*g)_{ij}(x,r)-g_{ij}(x,r)
\nn
&=&\xi(x)\left(2- r\,\pa_{r}\right)g_{ij}(x,r)+\nabla_i\,\xi_j(x,r)+\nabla_j\,\xi_i(x,r)+{\cal O}(\xi^2,\xi_i^2)~,
\eea
where the tildes were dropped from the point $(x,r)$. The expression is the same to linear order in the diffeomorphism because the difference of metrics is already of linear order.\\
\\
{\bf Proposition 1 (Infinitesimal version).} Let $\hat g$ be a Poincar\'e metric for $(M,[g])$ in normal form. If $\f:\,{\cal U}\rightarrow\,{\cal U}$ is invisible relative to $(\hat g,g)$ and reduces to $\varphi_M$, then $\vf_M$ is a conformal transformation.

To prove this, we take the expression \eq{dg} which was obtained from requirement that $\f$ be invisible relative to $\hat g$ in \eq{ij}-\eq{rr}. Requiring that $\f$ be invisible relative to $g$ as well, instructs us to set $(\f^* g)(p)= g(p)$, which is setting $\d_{\f^{-1}}\, g_{ij}(x,0)=0$. Thus, setting $r=0$ in \eq{dg}, this reduces to:
\bea
\d_{\f^{-1}}\, g_{ij}(x)&=&2\xi(x)\,g_{ij}(x)+\nabla_{i}\,\xi_{j}(x)+\nabla_{j}\,\xi_{i}(x)+{\cal O}(\xi^2,\,\xi_i^2)\nn
&=&2\xi(x)\,g_{ij}(x)+({\cal L}_\xi g)_{ij}(x)+{\cal O}(\xi^2,\xi_i^2)=0~,
\eea
where $\xi_i(x):=\xi_i(x,0)$ and ${\cal L}_\xi g$ is the Lie derivative with respect to the vector field $\xi$ on $M$ (not to be confused with the scalar function $\xi(x)$). 
Taking the trace of the above equation, and substituting the result back into the same equation, we get:
\bea
\xi(x)&=&-{1\over d}\,\nabla^i\,\xi_i(x)\label{xi}+{\cal O}(\xi^2,\,\xi_i^2)\\
\d_{\f^{-1}} g_{ij}(x)&=&{\cal L}_\xi\,g_{ij}(x)-{2\over d}\,g_{ij}(x)\,\nabla^k\,\xi_k+{\cal O}(\xi^2,\,\xi_i^2)=0~.\label{dg0}
\eea
This is precisely the conformal Killing equation, i.e.~the infinitesimal version of the condition for $\vf_M^{-1}$ (and hence $\vf_M$) to be a conformal transformation. $\Box$

As discussed in Section \ref{results}, the Killing equation \eq{dg0} on $M$ has, hitherto, been derived only in very special cases such as pure AdS space (cf.~Gubser et al.~(1998:~Eq.~(18))). The reason is that the more general treatments, like Imbimbo et al.~(2000:~Eq.~(2.6)) and Skenderis (2001:~Eq.~(8)), assume that $\xi_i(x)=0$, and hence they cannot get the Killing equation.

Equation \eq{dg0} can be rearranged as follows:
\bea\label{killing}
\d_{\varphi_M^{-1}}\,g(x)={\cal L}_\xi \,g(x)=-{2\over d}\,\nabla^k\xi_k(x)~g(x)~,
\eea
which is indeed the infinitesimal version of the following exponential form:
\bea\label{conftrafo}
(\varphi^{-1}_M)^*\,g(x)=e^{-2\xi(x)}\,g(x)~.
\eea
We will give a proof of this formula for finite diffeomorphisms at the end of this subsection.\\
\\
{\bf Proposition 2 (Infinitesimal version).} Let $\hat g$ be a Poincar\'e metric for $(M,[g])$ in normal form. If $\f:\,{\cal U}\rightarrow\,{\cal U}$ is invisible relative to $(\hat g,M)$, then $\f$ reduces to a Weyl transformation.

To prove this, notice that the requirement of invisibility relative to $M$ means that we have to set $\xi^i(x,0)=0$. But then we automatically get, from the requirement \eq{dg} that $\f$ be invisible relative to $\hat g$, that $\d_{\f^{-1}}\,g_{ij}(x,r)|_{r=0}=2\xi(x)\,g_{ij}(x)=(\f^*g)_{ij}(x)-g_{ij}(x)$. This is indeed an infinitesimal Weyl transformation. $\Box$

The finite version of the above is:
\bea\label{conftr}
\f^*g=e^{-2\xi(x)}\,g~.
\eea
This is the kind of Weyl transformation obtained in the standard accounts, see e.g.~Skenderis (2001:~Eq.~(10)): it is generated by the scalar $\xi(x)$, assuming that $\xi^i(x,0)=0$.\\
\\
{\bf Theorem 3.} Let $\hat g$ be a Poincar\'e metric for $(M,[g])$ in normal form. If $\f:\,{\cal U}\rightarrow\,{\cal U}$ is invisible relative to $(\hat g,M,g)$, then $\f$ is the identity.

If $\f$ is invisible relative to $M$ then $\xi_i(x)=0$, as we saw in Proposition 2. But since it is also invisible relative to $g$ then also $\xi(x)=0$, from \eq{xi}. Since \eq{r'} was valid over the entire $\psi(\,{\cal U})$, then $r=\ti r$ over the entire $\psi(\,{\cal U})$. 

In order to show that $\f$ is the identity, since we already know that $\xi^i(x,0)=0$, it is enough to show that the first derivative of $\xi^i(x,r)$ vanishes everywhere on $\,{\cal U}$. This now readily follows from \eq{ir'} because the right-hand side now identically vanishes. $\Box$

\subsubsection{Finite diffeomorphisms}

Let $\hat g$ be a Poincar\'e metric for $(M,[g])$ in normal form. Let $\f:\,{\cal U}\rightarrow\,{\cal U}$ be a finite diffeomorphism, invisible relative to $\hat g$. We use the same notation as before:
\bea
r&=&\ti r~\o(\ti x,\ti r)\nn
x^i&=&\ti x^i+\xi^i(\ti x,\ti r)~.
\eea
The generalisations of \eq{ij}-\eq{rr} in terms of these variables are as follows:
\bea\label{ijirrr}
{\pa x^k\over\pa\ti x^i}{\pa x^l\over\pa\ti x^j}~g_{kl}\left(x(\ti x,\ti r),r(\ti x,\ti r)\right)+\ti r^2\,\pa_i\o(\ti x,\ti r)\pa_j\o(\ti x,\ti r)&=&\o^2(\ti x,\ti r)~\ti g_{ij}(\ti x, \ti r)\nn
\pa_{\ti r} \xi^k~{\pa x^l\over\pa\ti x^i}\,g_{kl}\left(x(\ti x,\ti r),r(\ti x,\ti r)\right)+\half\,\ti r\,\ti\pa_i\left(\o^2(\ti x,\ti r)\right)&=&0\nn
\pa_{\ti r}\,\xi^i\pa_{\ti r}\xi^j\,g_{ij}\left(x(\ti x,\ti r),r(\ti x,\ti r)\right)+\ti r^2\,(\pa_{\ti r}\o)^2+\ti r\,\pa_{\ti r}(\o^2(\ti x,\ti r))&=&0~,
\eea
(Setting $\ti r=0$, the last equation implies that, if the metric is Riemannian rather than pseudo-Riemannian, then $\pa_{\ti r}\xi^i|_{\ti r}=0$. The same requrement is obtained for pseudo-Riemannian metrics from the requirement that the induced metric does not change: see \S\ref{invisible}. But we will not need this.)

Let us now assume that $\hat g$ is invisible relative to $g$ as well. Invisibility relative to $g$ gives:
\bea\label{ijirrr2}
{\pa x^k\over\pa\ti x^i}{\pa x^l\over\pa\ti x^j}~g_{kl}(x(\ti x))=\o^2(\ti x)~g_{ij}(x(\ti x))~,
\eea
where $\o(\ti x):=\o(\ti x,0)$. This is the analog of Proposition 2: the diffeomorphisms reduce to a boundary Weyl transformation.

Finally, if, in addition, $\hat g$ is invisible relative to $M$, so $x^i|_{\ti r=0}=\ti x^i$, then it follows that $\o(x)=+1$ (the plus sign chosen so as to preserve the orientation). That is, if the diffeomorphism along $M$ is the identity, then also the diffeomorphisms along the normal direction are the identity. This is the generalisation of Proposition 1. 


\subsection{Two classes of weakly-gravity invisible diffeomorphisms}\label{compare}

In this Section, I will compare the weakly-gravity invisible diffeomorphisms, obtained in Section \ref{invisibility}, to the physics literature.\footnote{I thank an anonymous referee for suggesting to make this comparison.}

The weakly gravity-invisible diffeomorphisms comprised two distinct classes: on the one hand, the diffeomorphisms invisible relative to $(\hat g,g)$, i.e.~satisfying (i) and (iii); on the other, the ones invisible relative to $(\hat g,M)$, i.e.~satisfing (i) and (ii). The former class gave rise to conformal transformations of the boundary manifold, i.e.~coordinate transformations at the boundary, satisfying the Killing equation \eq{killing}. The latter class gave rise to Weyl transformations, i.e.~local rescalings of the metric of the boundary manifold. 

These two classes are of course different, as diffeomorphisms of the metric $\hat g$: even if their effects, on the metric $g$ induced on the boundary, are similar---they both give rise to a local rescaling of the metric. The two classes are conceptually distinct: the former class is a coordinate transformation of the boundary manifold, whereas the latter class is a choice of a different representative of the conformal class of the metric. I now compare these two classes to the physics literature.

Diffeomorphisms of the former class, i.e.~invisible relative to $(\hat g,g)$, are, to {\it lowest} order, of the type (cf.~Proposition 1 in \S\ref{inficase}):
\bea\label{gkp}
x^i&=&\ti x^i+\xi^i(x)\nn
r&=&\ti r\left(1-\xi(x)\right),
\eea
where $\xi(x)=-{1\over d}\,\nabla^i\xi_i(x)$, and $\xi^i(x)$ satisfies the Killing equation ${\cal L}_\xi\,g_{ij}(x)={2\over d}\,g_{ij}(x)\,\nabla^i\xi_k$. Thus they correspond to conformal transformations at the boundary, i.e.~coordinate transformations of the boundary manifold which give rise to Weyl transformations of the metric. The Killing equation is the necessary and sufficient condition that they be Weyl transformations.

When restricted to pure AdS, this class is identical with the diffeomorphisms investigated in Gubser et al.~(1998:~\S2.1). These authors use the notation $z$ for my $r$, $\z^\m$ for my $\xi^i(\ti x,\ti r)$, and $\xi^\m$ for my $\xi^i(\ti x)$. Their $\xi^z$ corresponds to my $\xi(x)$. One easily verifies that their Eq.~(16) corresponds to my Eqs.~\eq{ir'} and \eq{xi}.

Diffeomorphisms of the latter class, i.e.~invisible relative to $(\hat g,M)$, are, to {\it lowest} order, of the type (cf.~Proposition 2 in \S\ref{inficase}):
\bea\label{imb}
x^i&=&\ti x^i\nn
r&=&\ti r\left(1-\xi(x)\right),
\eea
where now $\xi(x)$ is an {\it arbitary} smooth function, and there are no diffeomorphisms tangent to the boundary.\footnote{There are no $r$-independent diffeomorphisms, in other words, $\xi^i(x)=0$; but there are corrections at order $r^2$, if $\xi(x)$ is non-zero, i.e.~$\xi^i(x,r)\not=0$: cf.~Skenderis (2001:~Eq.~(10)).}

This class of diffeomorphisms corresponds to the one in Imbimbo et al.~(1999:~\S2). These authors use the coordinate $\r$ in Eq.~\eq{FG}, rather than the coordinate $r$ I have used in Eq.~\eq{NF} and in Section \ref{invisibility}.\footnote{Also, their metric induced at the boundary, $g_{ij}(x,\r)$, is rescaled by a factor of $\ell$ with respect to mine, so that their metric on $M$ has dimensions of length.} The change of coordinates is given by $\r=r^2/\ell$. One then easily checks that their Eq.~(2.2), with their choice of boundary condition $a^i(x,\r=0)=0$, is exactly Eq.~\eq{imb}. And it is in fact this choice of boundary condition that prevents them to finding the diffeomorphisms corresponding to Eq.~\eq{gkp} and the Killing equation. 

The difference between the two cases is the structures they preserve. The first class preserves $\hat g$ and $g$, i.e.~in particular, $\delta_{\f^{-1}}g_{ij}(x)=0$. For pure AdS, this amounts to considering bulk diffeomorphisms that leave the flat boundary metric (Euclidean or Minkowski) fixed. This means that the Weyl rescalings of the boundary metric and the coordinate transformations along the boundary directions must cancel each other out. This is the case for Eq.~\eq{imb}, under the conditions stated. The condition for the second class is that it preserves $\hat g$ and $M$, and the latter condition sets the components of the diffeomorphisms parallel to the boundary to zero, i.e.~$\xi^i(x,0)=0$. However, Weyl transformations are still allowed.

It is not surprising that the two classes of diffeomorphisms, Eqs.~\eq{gkp} and \eq{imb}, are only connected at the identity: since they are defined by the different structures that they preserve. By `connected at the identity', I here mean that one cannot simply set $\xi^i(x)=0$ in Eq.~\eq{gkp} to get Eq.~\eq{imb}, because then also $\xi(x)=0$, and then the diffeomorphism is the identity. This is of course the content of Theorem 3. 

\subsection{Gravity-invisible diffeomorphisms exist}\label{invisible}

In Theorem 3 of Section \ref{invisibility}, we proved that there are no strongly gravity-invisible diffeomorphisms close to the identity (i.e.~infinitesimal).\footnote{The arguments of \S\ref{invisibility} suggest that Propositions 1 and 2 (on $\f$ reducing to a conformal, respectively Weyl, transformation under various conditions) generalize to the finite case. A completely general proof of Theorem 3 in the finite case is likely to be possible, but it requires more work.} The strongly gravity-invisible diffeomorphisms form a natural class to consider because, though they preserve the normal form of the metric, they are not isometries of the $(d+1)$-dimensional metric: they are only isometries of the boundary conformal structure. Notice that the normal form of the metric corresponds to what physicists call a `radial gauge', i.e.~a choice of coordinates such that $\hat g_{ir}=0$. Thus, the strongly gravity-invisible diffeomorphisms preserve this gauge condition in addition to the two other invisibility conditions. We have shown that this class is trivial.

In this subsection we study the non-trivial class of gravity-invisible diffeomorphisms: those that are invisible relative to $(M,g)$. In the next section I will comment on the holographic interpretation of these gravity-invisible diffeomorphisms, as giving rise to QFT-invisible diffeomorphisms.

Our starting point is to rewrite the diffeomorphism in a form similar to \eq{xxtilde}:
\bea\label{rx}
r&=&\ti r-\xi(\ti x,\ti r)\\
x^i&=&\ti x^i+\xi^i(\ti x,\ti r)~.\nonumber
\eea

In order for $\f|_{M\times\{0\}}=1$, we must preserve the boundary $r=0$, i.e.~we must take\footnote{Because $\xi(\ti x,\ti r)$ in \eq{rx}, unlike \eq{xxtilde}, is not multiplied by $\ti r$, we have shifted the value of $\a$ up by one, i.e.~in \S\ref{inficase} we quoted the condition $\a\geq0$ for a diffeomorphism fixing the boundary: and this same condition is now stated as $\a\geq1$.} $\xi(\ti x,\ti r)=\ti r^\a\,\xi(\ti x)+{\cal O}(\ti r^{\a+1})$, $\xi^i(\ti x,\ti r)=\ti r^\b\,\xi^i(\ti x)+{\cal O}(\ti r^{\b+1})$, with $\a\geq 1$ and $\b\geq0$. I work to linear order in $\xi,\xi^i$ throughout.
The metric $\hat g$ in \eq{NF} is modified as follows:
\bea\label{modifiedg}
(\f^*\hat g)_{ij}&=&{\ell^2\over\ti r^2}\left(\left(1+\xi(\ti x,\ti r)\,\left({2\over\ti r}-\pa_{\ti r}\right)\right)\,g_{ij}(\ti x,\ti r)+\nabla_i\xi_j+\nabla_j\xi_i\right)\nn
(\f^*\hat g)_{ir}&=&{\ell^2\over\ti r^2}\left(-\pa_i\xi(\ti x, \ti r)+g_{ik}(\ti x,\ti r)\,\pa_{\ti r}\xi^k(\ti x,\ti r)\right)\nn
(\f^*\hat g)_{rr}&=&{\ell^2\over\ti r^2}\left(1-2\,\pa_{\ti r}\xi(\ti x,\ti r)+{2\over\ti r}\,\xi(\ti x,\ti r)\right)
\eea
where $\xi_i:=g_{ij}(\ti x,\ti r)\,\xi^j(\ti x, \ti r)$, and the covariant derivatives are with respect to the metric $g(\ti x,\ti r)$. Of course, if $\a=1$ and $\b=0$, the first formula agrees with the earlier result \eq{ij} and \eq{gtilde} when $\xi$, $\xi^i$ are expanded in $\ti r$. 

The gravity-invisible diffeomorphisms are only invisible relative to $(M,g)$ not the metric $\hat g$ on $\hat M$. So, we only need to demand that $\f$ is an isometry of the induced metric, obtained from the first of \eq{modifiedg} multiplying by a factor of $r^2/\ell^2$. We obtain the condition, at $r=0$:
\bea
g_{ij}(x,r)|_{r=0}=
\left(1-\xi(\ti x,\ti r)\,\partial_{\ti r}\right)g_{ij}(\ti x,\ti r)|_{r=0}+\nabla_i\,\xi_j(\ti x,\ti r)|_{r=0}+\nabla_j\,\xi_i(\ti x,\ti r)|_{r=0}~.
\eea
Let us now set $\xi(\ti x,\ti r)=\ti r^\a\,\xi(\ti x)$, $\xi^i(\ti x,\ti r)=\ti r^\b\,\xi^i(\ti x)$, set $r=0$, and use the fact that the first derivative of the metric is zero at lowest order in $r$.
\begin{figure}
\begin{center}
\includegraphics[height=3.2cm]{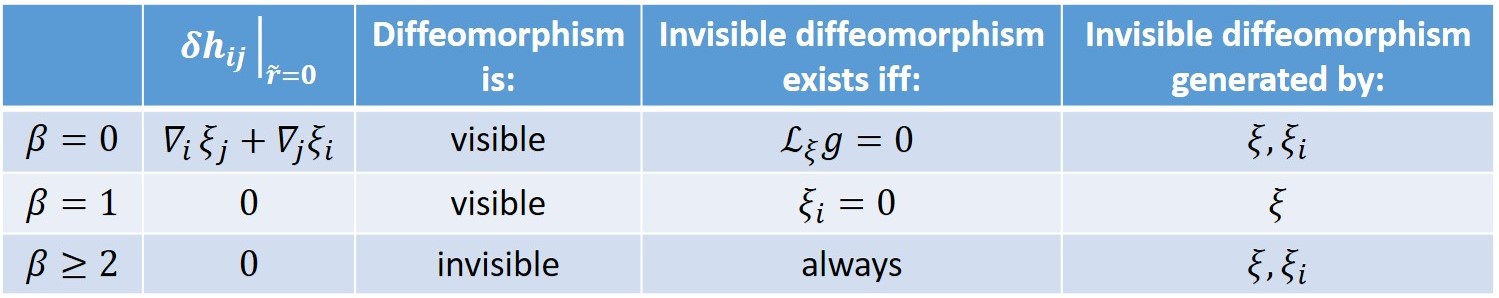}
\caption{\small{Table showing when an invisible diffeomorphism exists.}}
\label{Table}
\end{center}
\end{figure}

For $\b=0$, we find that the diffeomorphism is invisible unless 
\bea\label{naix}
\na_i\,\xi_j(\ti x)+\na_j\,\xi_i(\ti x)=0~,
\eea
i.e.~unless $\xi_i(\ti x)$ is an isometry of the representative of the boundary conformal structure $g$. For $\b\geq1$, we find that the diffeomorphism is always invisible. 

Let us consider a slightly stronger invisibility condition, namely that $(\f^*\hat g)_{ir}$, up to its conformal factor, should remain zero at $r=0$. This corresponds to the normal form of the metric (i) being preserved {\it asymptotically}. This requirement gives us the additional condition that $\xi_i(\ti x)=0$ when $\b=1$ in order to have an invisible diffeomorphism. The results are summarised in the table in Figure \ref{Table}. However this will not encumber the exposition in what follows. Since the additional condition is minimal, for it does not affect the other values of $\b$, when I discuss the physics of gauge-gravity dualities, I will still use $g$ instead of $\g$ for the induced metric, and will refer to the gravity-invisible diffeomorphisms as those that are invisible relative to $(M,g)$. 

In the gravity literature, the induced metric on any $d$-dimensional timelike hypersurface inside a $(d+1)$-dimensional volume is defined as: $\g_{\m\n}:=g_{\m\n}-n_\m n_\n$, where $n_\m$ is a normal covector to the hypersurface (see e.g.~Wald (1984, p.~255) for the spacelike case). Of course, this metric and $\hat g$ both give rise to the same induced metric $r=0$, and they give exactly the same invisibility conditions that we just obtained. This is shown in Appendix \ref{appendix}. 

In summary, there is an invisible diffeomorphism $\f$ (relative to $M$ and $g$) for $\b=0,1$ if ${\cal L}_\xi \,g=0$, resp.~$\xi_i=0$. This diffeomorphism is then generated by $\xi(\ti x)$. For $\b\geq2$, there is an invisible diffeomorphism (relative to $M$ and $g$) for any smooth $\xi$, $\xi^i$. See the table in Figure \ref{Table}.

\section{Invisibility in Gauge-Gravity Dualities}\label{ginvis}

In the previous Section, I derived two propositions and a theorem amounting to points (a)-(c) in Section \ref{results}. These results led to the definition, in \S\ref{invisible}, of gravity-invisible diffeomorphisms as those diffeomorphisms which are invisible relative to $(M,g)$. In this Section, I turn to the physical relevance of gravity-visible and gravity-invisible diffeomorphisms for gauge-gravity dualities.\footnote{The discussion in this Section and the next is adapted to the physics of interest. Therefore, the level of mathematical rigour will differ from that in the previous Section, though the results proven in Section \ref{visin} will be crucial in what follows.} For an introduction to gauge-gravity dualities, see Ammon and Erdmenger (2015). A conceptual introduction is in De Haro et al.~(2016a). 

The important question for gravity-visible and gravity-invisible diffeomorphisms, discussed in the previous Section, in connection with dualities, is whether they are also visible or invisible to the QFTs which are dual to the relevant gravity theories. To answer this question, we first need to discuss what the relevant gravity theory is. The definitions of invisibility in \S\ref{invisibility}, from which the propositions about gravity-invisibility and weak gravity-invisibility were derived, involve Poincar\'e metrics, which satisfy Einstein's equations in vacuum with a negative cosmological constant, Eq.~\eq{Einstein}, up to a specified order of approximation. The 1-parameter family of metrics $g_r$ on $M$ in \eq{NF} has an expansion of the form (Fefferman and Graham (2012:~Theorem 4.8):
\bea\label{formals}
g_r=\sum_{N=0}^\infty g_{r}^{(N)}\left(r^d\log r\right)^N~,
\eea
where each of the $g_{r}^{(N)}$ is a smooth family of metrics on $M$ even in $r$. Of particular interest is the term $N=0$, with its even expansion in $r$:
\bea\label{expansiongr}
g_r^{(0)}=g_{(0)}+r^2\,g_{(2)}+r^4\,g_{(4)}+\ldots,
\eea
and it follows from \eq{formals} that $g_{(0)}=g_0=g$.

For odd $d$, all $g_{r}^{(N)}$ with $N\geq1$ vanish, and only the $N=0$ term contributes: the above is then an even power series around $r=0$, and the solution is determined uniquely to infinite order given $g_{(0)}$ and $g_{(d)}$. Namely:
\begin{itemize}
\item All $g_{(n)}$ are determined algebraically from Einstein's equations (except for $g_{(0)}$ and $g_{(d)}$): they are given by covariant expressions involving $g_{(0)}$ and $g_{(d)}$ and their derivatives.
\item The coefficients $g_{(0)}$ and $g_{(d)}$  are not determined by Einstein's equations (only the trace and divergence of $g_{(d)}$ are determined): they are initial data.
\item One recovers pure AdS when $g_{(0)}$ is chosen to be flat (i.e. a flat Minkowski metric). In that case, all higher coefficients in the series \eq{expansiongr} vanish.
\end{itemize}

For even $d$, the logarithmic terms are nonzero, but again the entire $g_r$ is determined to infinite order given the same two data. In this expansion, Einstein's equations become algebraic equations relating the coefficients in the expansion \eq{expansiongr} for $g_r^{(N)}$ to $g_{(0)}$, $g_{(d)}$ and their derivatives. 

Thus, the discussion in Section \ref{visin} is relevant for the asymptotic solutions of Einstein's equations near the boundary of an Einstein space,\footnote{Notice that, for even $d$, the logarithmic terms only set in at order $r^d$.} where the metric induced on the boundary is arbitrary. Furthermore, the formal series \eq{formals} converges if the boundary conditions $g_{(0)}$ and $g_{(d)}$ are real-analytic functions of the boundary coordinates $x$ (Fefferman and Graham (2012:~p.~4, 49)).

As is well-known in the gauge-gravity literature (see e.g.~De Haro et al.~(2001:~Section 1), general relativity is a good approximation to the full string- or M-theory near $r=0$. As $r$ increases, new terms may be needed in the action (see the discussion in \S\ref{invisibleq}). In particular, all the asymptotic expansions used in the previous section (Eqs.~\eq{xxtilde}, \eq{r}, as well as the formulas evaluated at $r=0$) are good approximations as long as the size of the neighbourhood $\,{\cal U}\!$ is much smaller than the scale set by the radius of curvature. This means that the techniques developed here indeed give good approximations to the quantities of interest in gauge-gravity dualities, such as the holographic stress-energy tensor. 

In other words, the visible and invisible diffeomorphisms discussed in Section \ref{visin} are indeed relevant to the gravity side of gauge-gravity dualities. In \S\ref{invisibleq}, I will argue that gravity-invisible diffeomorphisms are invisible to the dual QFT, so they are also `QFT-invisible', and `duality-invisible'. In two Appendices \ref{examples} and \ref{matter}, I give explicit physics examples of the kinds of QFT quantities which are invisible, and discuss generalisations of these concepts to the case including bulk matter. In \S\ref{visibled}, I discuss how the weakly gravity-invisible diffeomorphisms are seen by the QFT, hence are QFT-visible.

\subsection{Gravity-invisible diffeomorphisms are QFT-invisible}\label{invisibleq}

In this subsection, I will discuss the sense in which gravity-invisible diffeomorphisms leave the gauge theory invariant, hence are `QFT-invisible'. On the conception of duality expounded in De Haro, Teh, and Butterfield (2016:~\S3.3), in order for invisible diffeomorphisms to be `gauge' in the philosophers' sense, they should leave all the quantities of the theory, evaluated on the states, invariant. As discussed, our theory is a theory of pure gravity: so our task now is to define the quantities that need to be evaluated.

The gauge-gravity duality isomorphism is often called a `dictionary', because it `translates' gravity to QFT quantities, and viceversa. This dictionary identifies the renormalized classical action with the generating functional of the QFT (in a suitably taken 't Hooft limit, see Ammon and Erdmenger (2015:~pp.~180-182)). For a theory of pure gravity without matter, the renormalized action is a functional (as usual, indicated by square brackets) of the representative $g$ of the boundary conformal structure, and nothing else  (for more details, see \S\ref{visibled}):
\bea\label{action}
S_{\tn{ren}}[g]\equiv W_{\tn{QFT}}[g]~.
\eea
In such a gravity theory with a boundary at spatial infinity, the basic classical physical quantity is the renormalised quasi-local stress-energy tensor $\Pi$ (Brown and York (1993), De Haro et al.~(2001:~Section 1)), which is evaluated by taking the derivative of \eq{action} with respect to the representative $g$ of the boundary conformal structure. But by the dictionary, (the one-point function of) this stress-energy tensor is precisely the 1-point function of the renormalized stress-energy tensor of the dual QFT at the fixed point, evaluated from the generating functional $W_{\tn{QFT}}[g]$! (for more details, see \S6.1.2 of De Haro et al.~(2016a)). That is:
\bea\label{stressenergy}
\bra\Pi_{ij}\ket_g\equiv\bra T_{ij}[g](x)\ket_{\tn{QFT}}={2\over\sqrt{g}}\,{\d W_{\tn{QFT}}[g]\over\d g^{ij}}={d\,\ell^{d-1}\over16\pi G_{\tn{N}}}\,g_{(d)ij}(x)+\mbox{(local terms)}~,
\eea
and the subscript $g$ indicates the fact that we are evaluating this expression on a state determined by the conformal class $[g]$ at the boundary. The first term on right-hand side of the above equation is the term appearing at order $r^d$ in the asymptotic expansion of the metric $g(x,r)$ \eq{expansiongr} in powers of $r$. The local terms are given in De Haro et al.~(2001:~Eq.~(1.3) and subsequent discussion). 

A calculation like \eq{stressenergy} can be done for CFTs whose only gauge invariant, local, renormalizable operators are built from their stress-energy tensor $T[g]$ only, for various representatives $g$ of conformal classes. By the state-operator correspondence which holds at the fixed point (cf.~e.g.~De Haro et al.~(2016a:~\S3)), further states are obtained by multiplying the reference state corresponding to $[g]$ with further powers  (even exponentials) of the stress-energy tensor.\footnote{There is no claim here that this exhausts the quantities in the CFT. Non-local quantities such as Wilson loops may also be required.}

Will the gravity-invisible diffeomorphisms preserve the 1-point function \eq{stressenergy}? Since $W_{\tn{QFT}}[g]$ is a functional of $g$, it must, in fact, be invariant under them. Of course, the stress-energy tensor is known to be anomalous, for even values of $d$, under conformal transformations of $g$, so that there is a dependence of the representative of the class chosen.\footnote{The anomaly was shown to be a consequence of the distinct behaviour between even and odd $d$ in the Poincar\'e metric considered in \S\ref{PNF}.} This will be our concern in \S\ref{visibled}. But, for the {\it gravity-invisible} diffeomorphisms that we are considering here  (cf.~Section \ref{results}, ``fix (ii)'' and ``fix (iii)''), $g$ is simply {\it invariant}, and therefore so is the generating functional.

There is an important question here, which relates to the passage from Horowitz and Polchinski (2006:~\S1.3.2) quoted at the end of the Introduction: `the gauge theory variables... are trivially invariant under the bulk diffeomorphisms, which are entirely invisible in the gauge theory.' Does my argument amount to saying that the generating functional $W_{\tn{QFT}}[g]$ is trivially invariant under the gravity diffeomorphisms? I submit that it does not amount to that. One important aspect of the non-triviality will appear when we discuss the higher-point functions in the paragraphs below---the dependence on the gravity metric is highly non-trivial: for the invisibility argument requires ensuring that $g_{(d)}$ is a functional of $[g]$ and $[g]$ only, i.e.~it requires having a global solution, of which not many are known (some examples will be given in Appendix \ref{examples}). But leaving this issue aside for the moment, and more importantly for the identification of the (simpliciter) gravity-invisible diffeomorphisms as QFT-invisible diffeomorphisms, we must ask: to what extent does the above `invisibility argument' about $W_{\tn{QFT}}[g]$ require {\it both} conditions (ii) and (iii) in \S\ref{invisibility} to obtain? Could we enlarge the class of QFT-invisible diffeomorphisms to contain the whole of (ii) or the whole of (iii) (or even their union!), rather than their {\it intersection}? Naively, one might be inclined to think that this is possible because both conditions should leave $g$ invariant: (ii) is the condition $\phi|_{M\times\{0\}}=0$, which in particular implies $\xi^i(\ti x,0)=0$ (where $\xi^i(\ti x,\ti r)$ is defined in Eq.~\eq{xxtilde}), and this means that there are no coordinate transformations on the boundary being induced. As for (iii), this is the condition $\f^*g=g$, which implies that $\d_{\f^{-1}}g_{ij}=2\xi\,g_{ij}+\nabla_i\,\xi_j+\nabla_j\,\xi_i=0$, and so the total effect on $g$ cancels out. 

But notice that the correct condition of QFT-invisibility is not that $g$ should not transform but rather that the diffeomorphism itself should be invisible to the QFT: it should not act on the QFT variables at all. In the case of (ii), setting $\xi^i(\ti x,0)=0$ still allows for $\f$ inducing a Weyl transformation on $g$, so that a diffeomorphism satisfying just (ii) is certainly QFT-visible. As for (iii), we still have non-trivial transformations $\xi,\xi_i$ which now jointly act on the boundary QFT variables. $\xi_i$ acts as a boundary coordinate transformation and $\xi$ acts as a Weyl transformation (rather than a conformal transformation) but just so that their combined effects cancel each other out. So, these diffeomorphisms {\it are} visible to the QFT, since they act on it as different transformations, even if, as a result of their combined effects, $g$ is left invariant under them. An additional reason {\it not} to classify diffeomorphisms fixing (iii) but not (ii) as QFT-invisible is that, if the generating functional $W_{\tn{QFT}}[g]$ depends on other (matter) couplings, the combined transformation of the matter couplings will not cancel out like they do for the metric, unless further transformations for the couplings are assumed---thus rendering the transformation, again, visible. In other words, (ii) and (iii) are both jointly needed if the diffeomorphisms are truly to qualify as QFT-invisible, rather than $g$ being `trivially invariant' under them. Thus, the (simpliciter) gravity-invisible diffeomorphisms are---in so far as the 1-point function is concerned---the correct candidates for QFT-invisible diffeomorphisms.

The argument extends to higher-point correlation functions of the stress-energy tensor:
\bea\label{general}
\bra T_{ij}(x_1)\cdots T_{kl}(x_n)\ket={2^n\over\sqrt{g(x_1)\cdots g(x_n)}}~{\d^{(n)}W[g]\over\d g^{ij}(x_1)\cdots \d g^{kl}(x_n)}~.
\eea
Now, when considering higher-point functions, the leading classical gravity approximation is valid when the underlying theory is string- or M-theory. Higher-order terms in the action will contribute corrections to the action, in the form of higher powers of the Riemann tensor and its derivatives, generically called `higher derivative terms': see e.g.~De Haro, Teh, and Butterfield (2016: ~\S4.1.2) for a discussion. Nevertheless, though the techniques of Section \ref{visin} which rely on the definition of a Poincar\'e metric do not apply to the general case including the higher derivative terms, the concepts of visibility and invisibility do apply, for the higher-derivative terms in the action are covariant: and, in so far as $W_{\tn{QFT}}[g]$ is a functional of $g$ only, the invisible diffeomorphisms will preserve the entire set of correlation functions \eq{general}. 

The functional $W_{\tn{QFT}}[g]$ is of course only known for very specific QFTs, typically defined on a space which is close to flat or under specific assumptions about the topology of the conformal structure.\footnote{There are further constraints on $W_{\tn{QFT}}[g]$ coming from the conservation law that applies to \eq{stressenergy}. For a discussion, see e.g.~Section 1.3 of van Rees (2010).} In Appendix \ref{examples}, I calculate this functional exactly, in the important case of four-dimensional self-dual gravity metrics. From the bulk point of view, the renormalized stress-energy tensor has to be calculated solution by solution, through the near-boundary expansion, as mentioned.

The higher-point functions are harder to calculate: for we would need to know the variation of $g_{(d)}$ (in \eq{stressenergy}) with respect to an {\it arbitrary} metric $g$, and, in the general case, this is beyond the reach of current techniques. However, we can compute it in specific cases, as I will illustrate in two examples, in Appendix \ref{examples}: of fluctuations around pure AdS, and of self-dual and massive gravity solutions.

In other words, checking that the correlation functions of the stress-energy tensor are indeed invisible to gravity-invisible diffeomorphisms requires the existence of a global solution: so that $g_{(d)}$ is indeed a functional of $g$. And, after all is said and done, the gravity-invisible diffeomorphisms defined in \S\ref{invisibility} do indeed come out as the correct QFT-invisible diffeomorphisms.

In the presence of matter in the bulk, the QFT at the fixed point has further operators, for instance $\bra T_{ij}(x_1)\,O(x_2)\ket$, where $O(x_2)$ is a local operator which can be constructed out of new fields, and accordingly we get more states. I will comment on this case in Appendix \ref{matter}. For a review of dualities for gravity coupled to matter, see Skenderis (2002).

\subsection{Weakly gravity-visible diffeomorphisms are QFT-visible}\label{visibled}

In the previous subsection, and in Appendix \ref{examples}, we studied simply invisible diffeomorphisms in some detail, and concluded that they indeed preserve the physical quantities of the QFT. Now I will discuss how the weakly gravity-invisible diffeomorphisms found in Section \ref{visin} are seen by the QFT, hence are `QFT-visible'. By Propositions 1 and 2, weakly gravity-invisible diffeomorphisms give rise to a conformal, respectively Weyl, transformation of the representative of the boundary conformal structure $g$, as in \eq{conftr}. There are two ways in which these diffeomorphisms are indeed visible to the QFT. 

The first way in which weakly-gravity invisible diffeomorphisms are visible to the QFT is their giving rise to conformal or Weyl transformations of the boundary QFT. More precisely, the boundary theory is a QFT at a conformally invariant fixed point, or a CFT. If the representative of the boundary conformal metric transforms under a weakly gravity-invisible diffeomorphism as $\vf^{-1}_M: g(x)\mapsto e^{-2\xi(x)}\,g(x)$ (see \eq{conftrafo}), then transforming the other fields $\F_i(x)$ in the CFT (where $i$ runs over the different species of fields) with specific weights $w_i\in\mathbb{R}$, $\F_i(x)\mapsto e^{w_i\,\xi(x)}\,\F_i(x)$, renders the theory (classically) invariant. But clearly, such a diffeomorphism is {\it visible} to the QFT: it is a conformal transformation of the fields. 

There is a second way in which weakly-gravity invisible diffeomorphisms are visible to the QFT. Conformal transformations constitute a classical symmetry of the QFT at the fixed point but they are not always a symmetry of the {\it quantum} theory. There is a conformal anomaly for even values of the boundary dimension $d$ (Henningson and Skenderis (1998)).\footnote{The following discussion follows the exposition in Skenderis (2000:~Section 3).} The gravity action is IR divergent due to the infinite volume of $\hat M$, as can be seen from the divergence of the Poincar\'e metric \eq{NF} at $r=0$, and so is renormalized in Eq.~\eq{action}. Thus the action needs to be regularised, introducing a cutoff $r=\e$, and renormalized (De Haro et al.~(2001:~Section 3)). For even $d$, the renormalization procedure breaks the covariance of the action: one of the counterterms that are needed introduces a dependence of the classical action on the chosen representative of the boundary conformal structure. So, the classical action is anomalous under such transformations:
\bea
S_{\tn{ren}}[e^{-2\xi(x)}\,g]=S_{\tn{ren}}[g]+{\cal A}\,[g,\xi]~,
\eea
where ${\cal A}$ is the anomaly, which, for infinitesimal $\xi$, was computed in Henningson and Skenderis (1998:~Section 3). Applying \eq{stressenergy} and using the identification \eq{action}, it now follows that the stress-energy tensor transforms, under such diffeomorphisms, as:
\bea\label{anomalous}
\bra T_{ij}[e^{-2\xi(x)}\,g](x)\ket=e^{(d-2)\,\xi(x)}\left(\bra T_{ij}[g](x)\ket+{1\over\sqrt{g}}\,{\d{\cal A}[g]\over\d g^{ij}(x)}\right).
\eea
This transfomation law can be found in De Haro et al.~(2001:~Appendix C), for infinitesimal $\xi$. Of course, if we take the trace of \eq{anomalous}, we reproduce the conformal anomaly, which was well-known in the conformal field theories in $d=2$, $d=4$, but had never been computed before in the $d=6$ theory that is dual to 7-dimensional Einstein gravity (see Deser and Schwimmer (1993), Henningson and Skenderis (1998:~Section 3): also De Haro et al.~(2016:~\S4.2.1)).

This shows that already the 1-point function of the stress-energy tensor exhibits anomalous behaviour under these diffeomorphisms, for even $d$: and, in this sense, the diffeomorphisms are visible and the theory is not conformally invariant. Notice that, in the QFT at the fixed point, this anomaly is a {\it quantum} effect, which is mirrored by the divergence of the classical gravity action.

\section{Discussion and Conclusions}\label{discussionandc}

In this paper I have presented a number of results which: (i) make rigorous a number of physical intuitions about asymptotic symmetries in general relativity with a negative cosmological constant, and in gauge-gravity dualities; (ii) provide the mathematical and physical basis for the philosophical comparison of duality and gauge symmetry presented in De Haro, Teh, and Butterfield (2016:~\S\S5-6); (iii) underpin the discussion of background-independence for gauge-gravity dualities in De Haro (2016:~\S\S 2.3.2-2.3.4). 

These results are of physical interest in their own right. While the general gist of some of them may be known to experts in the conformal geometry of gauge-gravity duality, the mathematical and conceptual details are novel: and they bear on physical and philosophical discussions of general relativity and of duality. 

As I have argued, the notion of weakly gravity-invisibility naturally makes precise the idea of asymptotic symmetries studied in the literature on general relativity and on gauge-gravity duality. These asymptotic symmetries are expected to induce the conformal transformations of the CFT.  In this paper, weakly gravity-invisibility is defined as the preservation of appropriate structure, viz.~the normal form of the metric and in addition either the boundary manifold $M$, in terms of its points (i.e.~the diffeomorphism goes to the identity at infinity), or the representative of the conformal class $g$. \\

The first result was that weakly gravity-invisible diffeomorphisms give QFT-visible diffeomorphisms, namely, they indeed induce the conformal transformations of the CFT. Furthermore, the class of weakly gravity-invisible diffeomorphisms was shown to be {\it larger} than normally expected. \\

Secondly, the class of non-trivial strongly gravity-invisible diffeomorphisms was found to be empty: the only strongly gravity-invisible diffeomorphism is the identity. This is a surprising new result, because these diffeomorphisms were defined as preserving a very general, normal, form of the metric (a class of Poincar\'e metrics), while in addition preserving the conformal manifold $M$ and the representative of the conformal class $g$. \\

Finally, the class of gravity-invisible diffeomorphisms turns out, as a consequence of the second point, smaller than expected. Nevertheless, non-trivial gravity-invisible diffeomorphisms do exist and are given by the diffeomorphisms satisfying both (ii) and (iii) in \S\ref{invisibility}. There is of course no claim here that gravity-invisible diffeomorphisms {\it exhaust} the QFT-invisible diffeomorphisms: for there could be other diffeomorphisms that fit the bill.\footnote{However, the larger class that results from dropping (ii) was argued not to be a good candidate for QFT-invisibility because such diffeomorphisms do act on the CFT states: even if, in certain cases, the combined effects cancel each other out.} The argument, in \S\ref{invisibleq}, that the gravity-invisible diffeomorphisms provide genuine QFT-invisible diffeomorphisms, in the sense that they do not act on the QFT, leaving all of its physical quantities unchanged, is rather non-trivial. \\

In the literature, the characterisation of the asymptotic symmetries is not always very precise. For instance, the class of non-trivial asymptotic symmetries---corresponding to my QFT-visible diffeomorphisms---is sometimes limited to only those diffeomorphisms which fix the radial direction $r$ (Janiszewski and Karch (2013:~\S1.2)). But this is too restrictive: for, as we saw in Propositions 1 and 2 (see also the comment at the end of \S\ref{invisibility}), the two classes of QFT-visible diffeomorphisms (those fixing (i) and (ii), and those fixing (i) and (iii)) have non-zero $\xi(x)$ (non-zero $\l(x)$, in the notation of the comment in \S\ref{invisibility}), which parametrises the change of the radial coordinate: they act nontrivially along the $r$-direction while fixing $r=0$. 

Also, it is sometimes claimed that the QFT-invisible diffeomorphisms are those that `go to unity at the boundary' (this being the class by which the allowed diffeomorphisms have to be quotented in order to obtain the asymptotic symmetry group). But also this is imprecise: for a diffeomorphism can go to unity at the boundary (i.e.~fixing (ii)) while still modifying the representative of the boundary conformal class through its $r$-dependence resulting in a rescaling of the metric $\xi(x)$, as shown in the proof of Proposition 2.

The correct QFT-invisibility condition to require is that the diffeomorphisms must fix both (ii) and (iii). Also, Theorem 3 ensures that the QFT-invisible diffeomorphisms form a class that is disjoint from the class of QFT-visible diffeomorphisms, i.e.~the triviality of the class of diffeomorphisms fixing all of (i), (ii), (iii) means that the intersection between the QFT-visible and the QFT-invisible is empty. Thus in my construction there is no need to quotient the QFT-visible diffeomorphisms, as defined in \S\ref{visibled}, by that putative intersection.\\

The construction of a clear notion of QFT-invisible diffeomorphisms for general relativity and for gauge-gravity dualities carried out here, underlies the philosophical comparison in De Haro, Teh, and Butterfield (2016:~\S2,~\S\S 5-6) between duality and gauge symmetry. More precisely, in that paper (\S2) a distinction was made between: (i) gauge symmetries which are (Redundant), i.e.~roughly: the formulation of the theory uses more variables than the number of degrees of freedom of the system being described; and (ii) gauge symmetries which are (Local), i.e.~spacetime-dependent transformations. While the diffeomorphisms considered in this paper are all (Local) in this sense, not all of them are (Redundant).\footnote{In the physics literature, what is here called (Redundant) is sometimes called a `gauge symmetry', while a transformation which is (Local) but not (Redundant) is sometimes called a `global' symmetry. This use of  `global' and `gauge' seems confusing because, as just mentioned, symmetries which are not (Redundant) can be (Local), hence they should not be called `global'. Therefore I adopt the characterisation of gauge symmetries given in De Haro et al.~(2016:~\S2).} 

Let us now discuss which diffeomorphisms are (Redundant). The QFT-invisible diffeomorphisms are (Redundant): because the physical quantities do not depend on them. 

On the other hand, the QFT-visible diffeomorphisms are potentially physical, in which case they cannot be (Redundant). In De Haro, Teh, and Butterfield (2016:~\S6), the analogous case of Galileo's ship thought experiment was used to illustrate how these diffeomorphisms, which are non-trivial at the boundary, can generate a relational physical difference between a proper subsystem and its environment when the action of the symmetry is restricted to the subsystem. Because of this characterization  as a `subsystem', these diffeomorphisms can indeed become physical. The condition for them to be physical can be cashed out in terms of what in De Haro (2016a:~\S1)\footnote{That paper builds on Dieks et al.~(2015:~\S3.3.2), and promotes the latter's `internal viewpoint' to an `internal interpretation'.} is called an `external interpretation'. Such an external interpretation indeed treats the world described by the theory as a subsystem. On an external interpretation, then, QFT-visible diffeomorphisms are (Local) but not (Redundant).

But if an `internal interpretation' is available: then, at least in the case of odd $d$---in which the conformal anomaly vanishes---the conformal symmetry might well be taken to be a redundancy of the theory's formulation. Thus in this case the QFT-visible diffeomorphisms would become (Redundant). The conditions for an internal interpretation to obtain are spelled out in De Haro (2016a:~\S1).

Having specified the class of QFT-invisible diffeomorphisms, this can now be used to formulate a hole argument, labelled a `bulk' argument, for Einstein spaces with a negative cosmological constant. \\

Most of the technical results presented in this paper rely only on the properties of the asymptotic behaviour of the Poincar\'e metric. Verifying that a diffeomorphism, for a given metric, is gravity-invisible involves just its asymptotic expansion. However, showing that the gravity-invisible diffeomorphisms are also QFT-invisible does involve assumptions about the global behaviour of the solutions, as we saw in \S\ref{invisibleq} and, especially, in Appendix \ref{examples}: where QFT correlation functions were computed using global solutions. 

Because the Fefferman-Graham expansion \eq{formals} turns the problem of solving a differential equation asymptotically into that of solving a set of coupled algebraic relations, the results that only depend on the asymptotic solutions can be analytically continued: from the case of a negative cosmological constant or AdS, to the case of a positive cosmological constant, or de Sitter space, $\ell_{\tn{AdS}}\mapsto i\ell_{\tn{dS}}$.\footnote{For an analysis directly in de Sitter space, see Anninos et al.~(2011).} Indeed, the bulk hole argument of De Haro et al.~(2016: Section 6) only strictly requires an infinitesimal neighbourhood of the boundary. Hence the gravity-invisible diffeomorphisms defined here can be used to make a bulk/hole argument for (generalised) de Sitter space, where the anti-de Sitter boundary is mapped to timelike future infinity in de Sitter space.

But the methods developed here, of constructing diffeomorphisms that preserve the relevant asymptotic structures, should readily generalize to cases in which there is no Fefferman-Graham structure like Eqs.~\eq{NF} and \eq{formals} but instead some other kind of asymptotic expansion. For one such class of examples, in which the QFT is a non-relativistic quantum field theory, i.e.~the spacetime has a globally defined time coordinate and a preferred foliation, see Janiszewski and Karch (2013:~\S1.2). In such a case, the QFT-visible diffeomorphisms should be the ones that preserve the corresponding structure, and which at the boundary induce the symmetries of the non-relativistic QFT.

The distinction between QFT-visible vs.~invisible diffeomorphisms establishes the two relevant kinds of diffeomorphisms which were discussed in detail in De Haro (2016:~\S2.3.3), and to which two kinds of analyses of background-independence applied. These two kinds of diffeomorphisms taken together formed what in that paper were called diffeomorphisms that `preserve the asymptotic form of the metric' (labelled (a1)). In this paper I have thus specified that the metric structure preserved can be either: (b: QFT-visible) the normal form of the metric and in addition $M$ or $g$; (c: QFT-invisible) $M$ and $g$. The mathematical details of the discussion (De Haro (2016:~\S2.3.3)) of the (lack of) covariance of the physical quantities for the QFT-visible diffeomorphisms, for even values of $d$, have now been fleshed out in \S\ref{visibled}, and in particular the anomalous transformation is in Eq.~\eq{anomalous}. In De Haro (2016:~\S2.3.3), the two kinds of diffeomorphisms (a1) discussed here were distinguished from yet another class, labelled (a2), of `large' diffeomorphisms: which do not preserve any of the pairwise structures (a)-(c) considered in the Introduction. These diffeomorphisms are expected to map solutions to inequivalent solutions. It would be interesting to investigate this class of diffeomorphisms.

These results also make it possible to now discuss the important philosophical question of the possible emergence of diffeomorphism invariance. 

\section*{Acknowledgements}
\addcontentsline{toc}{section}{Acknowledgements}

I thank Dionysios Anninos, Abhay Ashtekar, Daniel Grumiller, James Read, Kostas Skenderis, Nicholas Teh and, especially, Jeremy Butterfield for insightful discussions. I would also like to thank three anonymous referees, and several audiences: the workshop {\it L'\'emergence dans les sciences de la mati\`ere} in Louvain-la-Neuve; 
{\it Carlofest}, Carlo Rovelli's 60th birthday conference in Marseille; the Munich Center for Mathematical Philosophy, LSE's Sigma Club, and the Oxford philosophy of physics group.
This work was supported by the Tarner scholarship in Philosophy of Science and History of Ideas, held at Trinity College, Cambridge.

\appendix
\section{A Condition for Gravity-Invisibility}\label{appendix}

At the end of \S\ref{invisible}, I discussed an alternative definition the induced metric $\g_{\m\n}$, which is common in general relativity. In this appendix I compute the induced metric and show that it gives the same conditions for invisibility.

The normal covector to the boundary $r=0$, $n={\ell\over r}\,\dd r$, transforms as:
\bea
\f^*n&=&{\ell\over\ti r}\left(\left(1-\pa_{\ti r}\xi(\ti x,\ti r)+{1\over\ti r}\,\xi(\ti x,\ti r)\right)\dd r-\pa_i\xi(\ti x, \ti r)\,\dd x^i\right)~.
\eea
The induced metric on the boundary is: $\g_{\m\n}:=g_{\m\n}-n_\m n_\n$, so that it transforms as:
\bea
(\f^*\g)_{ij}&=&{\ell^2\over\ti r^2}\left(\left(1+\xi(\ti x,\ti r)\left({2\over\ti r}-\,\pa_{\ti r}\right)\right)g_{ij}(\ti x,\ti r)+\nabla_i\xi_j+\nabla_j\xi_i\right)\nn
(\f^*\g)_{ir}&=&{\ell^2\over\ti r^2}~g_{ij}(\ti x,\ti r)~\pa_{\ti r}\xi^j(\ti x,\ti r)\nn
(\f^*\g)_{rr}&=&0
\eea
Let us define the transformed metric $\hat h:=\phi^*\g=:{\ell^2\over r^2}\,h$, then the conformal metric is $h={r^2\over\ell^2}\,\f^*\g$. Its components are:
\bea
h_{ij}&=&\left(1-\xi(\ti x,\ti r)\,\pa_{\ti r}\right)g_{ij}(\ti x,\ti r)+\nabla_i\,\xi_j(\ti x,\ti r)+\nabla_j\,\xi_i(\ti x,\ti r)\nn
h_{ir}&=&g_{ij}(\ti x,\ti r)~\pa_{\ti r}\xi^j(\ti x,\ti r)~.
\eea

Since $\g$ is the metric normal to the vector $n^\m$, whose components are tangential to the boundary, the criterion for QFT-invisibility is that $\f$ leaves this metric unmodified. We will now calculate whether, for specific values of $\a$ and $\b$ above, there are any obstructions for the existence of invisible diffeomorphisms. As before, we set $\xi(\ti x,\ti r)=\ti r^\a\,\xi(\ti x)$, $\xi^i(\ti x,\ti r)=\ti r^\b\,\xi^i(\ti x)$. 

For $\b\geq2$, we get $h_{ij}|_{\ti r=0}=g_{ij}$ and $h_{ir}|_{\ti r=0}=0$ and such a diffeomorphism is invisible. 

For $\b=1$, the components along the boundary are still unaffected, but $h_{ir}|_{\ti r=0}=\xi_i$ and the diffeomorphism is visible, like before. 

For $\b=0$, we get $h_{ir}|_{\ti r=0}=0$ identically. But the diffeomorphism has a {\it visible} effect: it generates non-zero $h_{ij}|_{\ti r=0}=\nabla_i\xi_j+\nabla_j\xi_i$, as before.

\section{QFT Correlation Functions: Three Examples}\label{examples}

In this Appendix, I will calculate the correlation functions \eq{general} in three examples of bulk solutions, which will exhibit explicitly the dependence between $g_{(d)}$ and $g$, discussed in \S\ref{invisibleq}. The aim is to show the independence of the QFT quantities \eq{general} from the gravity-invisible diffeomorphisms, thus illustrating the general point made in \S\ref{invisibleq}.  These examples thus show that the mathematical theory developed in Section \ref{visin} is relevant in the important physical sense of being instantiated in non-trivial examples of gauge-gravity dualities.

I will consider solutions with Euclidean signature and $d=3$, i.e.~solutions of a four-dimensional, Euclidean theory in the bulk.

\subsection{Perturbations of a flat boundary}

Consider a representative of the boundary conformal structure that is almost flat, 
\bea
g_{ij}(x,r)=\d_{ij}+h_{ij}(r,x)~,
\eea
where $h_{ij}$ is the linealised fluctuation around the Euclidean solution. It is not hard to show that Einstein's equations have a unique regular, linearized solution, which can be written entirely in terms of the transverse, traceless part of $h_{ij}(r,x)$. (The solution is in Section 2 of De Haro~(2008)). This solution then has itself an expansion in $r$, the coefficients of which satisfy:
\bea
\bar h_{(3)}={1\over3}\,|\Box|^{3/2}\,\bar h_{(0)}~,
\eea
where $\bar h$ denotes the transverse, traceless part of $h$, obtained by projecting: $\bar h_{ij}=\Pi_{ijkl}\, h_{kl}$. Substituting this into \eq{stressenergy} thus gives the 1-point function of the stress-energy tensor, and the two-point function is obtained by a further variation (in momentum space):
\bea
\bra T_{ij}T_{kl}\ket={\ell^2\over8\pi G_{\tn{N}}}\,|p|^3\,\Pi_{ijkl}~.
\eea
This two-point function is indeed invariant under gravity-invisible transformations. In this particular case, all the higher-point functions vanish.

\subsection{Self-dual solutions}

Another case of interest is that of bulk solutions for which the Weyl tensor is self-dual:
\bea\label{SD}
C_{\m\n\a\b}=\half\,\e_{\m\n\l\s}\,C_{\a\b}{}^{\l\s}~,
\eea
and in the anti-self-dual case there is a relative minus sign. These solutions have a $g_{(3)}$, i.e.~a boundary condition which equals the Cotton tensor $C_{ij}$ (De Haro~(2008)), and so the stress-energy tensor is:
\bea\label{instanton}
\bra T_{ij}\ket={\ell^2\over8\pi G_{\tn{N}}}\,C_{ij}[g]~,
\eea
where the two-index Cotton tensor is a traceless and conserved 2-tensor,\footnote{The significance of the Cotton tensor is that it plays, in three dimensions, the role which the Weyl tensor (which vanishes identically in three dimensions) plays in dimensions higher than three: it is the tensor whose vanishing is a sufficient and necessary condition for conformal flatness.} given by:
\bea\label{Cotton}
C_{ij}:=\half\,\e_i{}^{kl}\,\nabla_k\left(R_{jl}-{1\over4}\,g_{jl}\,R\right)~.
\eea
I emphasize that the stress-energy tensor \eq{instanton} is exact: there is no linearization involved. The bulk solutions satisfying \eq{SD} are usually called gravitational instantons. Because the result \eq{instanton} is exact, we can (very exceptionally!) calculate the {\it exact} generating functional, up to a constant. It is given by the gravitational Chern-Simons action:
\bea
W_{\tn{QFT}}=-{\ell^2\over32\pi G_{\tn{N}}}\,\int\Tr\left(\G\wedge\dd\G+{2\over3}\,\G\wedge\G\wedge\G\right),
\eea
and an anti-self-dual solution has a plus sign. Furthermore, we can calulate from \eq{instanton} {\it all} the higher-point functions \eq{general} in the QFT, which correspond to self-dual solutions \eq{SD}. 

For instance, take a squashed 3-sphere, with metric:
\bea
g&=&{\ell^2\over4}(\s_1^2+\s_2^2+\a\,\s_3^2)\nn
\s_1+i\s_2&=&e^{-i\psi}\left(\dd\th+i\,\sin\th\,\dd\varphi\right)\nn
\s_3&=&\dd\psi+\cos\th\,\dd\varphi~,
\eea
where $\a$ is the squashing parameter. The round three-sphere is obtained when $\a=1$, and the scalar curvature is $R={8\over\ell^2}\left(1-{\a\over4}\right)$. As a function of the metric, the Cotton tensor is then given by the following identity: $\Ric-{1\over3}\,R\,g+{\ell\over3\sqrt{\a}}\,C=0$. Explicitly, it takes the following form:
\bea
C={1\over\ell}\left(\a-1\right)\sqrt{\a}\left(\s_1^2+\s_2^2-2\a\,\s_3^2\right).
\eea
Of course, for a round three-sphere, i.e.~$\a=1$, the Cotton tensor vanishes.


The mathematical interest in these solutions goes back to Pedersen and LeBrun..., but was revived by the Fefferman-Graham results in Anderson... For some examples of gravitational instantons, see e.g.~Martelli et al.~(2013).

\subsection{Topologically massive gravity}

The previous kind of solution can be generalised to so-called topologically massive gravity theories in three dimensions. The starting point of the generalization is to include,  in the boundary condition, in addition to the Cotton tensor, a Ricci term:
\bea
{3\ell^2\over\,32\pi G_{\tn{N}}}~g_{(3)}=\m\,\Ric[g]-C[g]~,
\eea
where $\m$ is the mass introduced, on dimensional grounds, by the boundary condition. Again, for such solutions the all-order \eq{general} can be computed. At the linearized level, the 2-point function is:
\bea
\bra T_{ij}T_{kl}\ket={\ell^2\over8\pi G_{\tn{N}}}\,|p|^3\,\Pi_{ijkl}+{ip^2\over\m}\,\e_{imp}\,p_n\,\Pi_{jmkl}~.
\eea

To conclude this Appendix: As claimed, these examples illustrate how: (i) the higher-point functions, which are the quantities of interest in the theory, can be calculated, and (ii) that they are invariant under invisible diffeomorphisms as defined in \S\ref{invisible}. Thus, the mathematical theory here developed is physically, as well as mathematically, non-empty.

\section{Coupling Gravity to Matter}\label{matter}

Showing that the methods of Fefferman and Graham generalise to theories with matter was the topic of Section 5 in De Haro et al.~(2001). In the same way that Poincar\'e metrics can be constructed with given boundary conformal data, matter fields (such as scalar fields satisfying the Klein-Gordon equation, or gauge fields satisfying their equation of motion) can be solved for by using similar methods, i.e.~solving an asymptotic series given boundary conditions on $M$. A similar structure arises, with the expectation values of an operator $\bra O_\D(x)\ket$ of scaling dimension $\D$ being given by the 1-point functions of the canonical momenta associated to bulk matter. Of course, now one has to solve the coupled gravity-matter equations: but this can be done asymptotically: and for fields within the unitarity bounds of the dual QFTs, the Fefferman-Graham expansion works. For illustration, the one-point function dual to a scalar field of mass $m$ is given by:
\bea
\bra O_{\D_+}(x)\ket=\left(2\D_+-d\right)\f_{(2\D_+-d)}+(\mbox{local terms})~,
\eea
where $\D_+$ the dimension of the operator and $\f_{2(\D_+-d)}$ is the coefficient in the Fefferman-Graham expansion at order $\D_+$:
\bea
\f(x,r)=r^{\D_-}\left(\f_{(0)}(x)+r\,\f_{(1)}(x)+\ldots+r^{\D_+-\D_-}\,\f_{(2\D_+-d)}\right),
\eea
where $\D_\pm:={d\over2}\pm\sqrt{{d^2\over4}+m^2\ell^2}$. $\D_+$ is the scaling dimension of the operator $O(x)$ in the QFT.

Now if the dimensions of operators that we add to the QFT are within the unitarity bounds, the back-reaction of the fields on the metric does not affect its leading behaviour (De Haro et al.~(2001:~\S5.2)). So, although the technical details in the derivation of the invisible diffeomorphisms in Section \ref{visin} will differ, the conclusion (c), in \S\ref{results}, that there is a non-empty class of such diffeomorphisms will be unaffected, and therefore the invisibility analysis in Section \ref{ginvis} is not affected by the addition of matter.

\section*{References}
\addcontentsline{toc}{section}{References}

Ammon, M., Erdmenger, J.~(2015). {\it Gauge/Gravity Duality. Foundations and Applications}. Cambridge University Press, Cambridge.\\
\\
Anninos, D., Ng, G.S., Strominger, A. (2011). `Asymptotic Symmetries and Charges in De Sitter Space', {\it Classical and Quantum Gravity} 28, 175019. doi:10.1088/0264-9381/28/17/175019   [arXiv:1009.4730 [gr-qc]].\\
\\
Arnowitt, R., Deser, S., and Misner, C.~W.~(1959). `Dynamical structure and definition of energy in general relativity'. {\it Physical Review}, 116(5), p.~1322.\\
\\
Arnowitt, R.~L., Deser, S., and Misner, C.~W.~(2008). `Republication of: The dynamics of general relativity'. {\it General Relativity and Gravitation}, 40(9), pp.~1997-2027.  doi:10.1007/s10714-008-0661-1
  [gr-qc/0405109].\\
\\
Ashtekar, A.~and Hansen, R.~O.~(1978). `A unified treatment of null and spatial infinity in general relativity. I. Universal structure, asymptotic symmetries, and conserved quantities at spatial infinity'. {\it Journal of Mathematical Physics} 19, pp.~1542-1566; doi: 10.1063/1.523863\\
\\
Ashtekar, A., Bonga, B., and Kesavan, A.~(2015). `Asymptotics with a positive cosmological constant: I. Basic framework', {\it Classical and Quantum Gravity}, 32, no.2,  p.~025004,
  doi:10.1088/0264-9381/32/2/025004
  [arXiv:1409.3816 [gr-qc]].\\
\\
Bondi, H., van der Burg, M.~G.~J., Metzner, A.~W.~K.~(1962). `Gravitational Waves in General Relativity. VII. Waves from Axi-Symmetric Isolated Systems', {\it Proceedings of the Royal Society of London. Series A, Mathematical and Physical Sciences}, 269 (1336), pp.~21-52.\\
\\
Brown, J.~D.~and Henneaux, M.~(1986). `Central Charges in the Canonical Realization of Asymptotic Symmetries: An Example from Three-Dimensional Gravity', {\it Communications in Mathematical Physics} 104, p.~207.\\
\\
Brown, J.D., York, Jr., J.W.~(1993). `Quasilocal energy and conserved charges derived from the gravitational action', {\it Physical Review D}, 47, pp.~1407-1419 
  [gr-qc/9209012].\\
\\
De Haro, S., Skenderis, K., and Solodukhin, S. (2001). `Holographic reconstruction of spacetime and renormalization in the AdS/CFT correspondence", \emph{Communications in Mathematical Physics}, 217 (3), pp.~595-622. [hep-th/0002230].\\
\\
De Haro, S.~(2008). `Dual Gravitons in AdS(4) / CFT(3) and the Holographic Cotton Tensor'. {\it Journal of High-Energy Physics}, 0901 (2009) 042.
  doi:10.1088/1126-6708/2009/01/042
  [arXiv:0808.2054 [hep-th]].\\
\\
De Haro, S. (2016). `Dualities and emergent gravity: Gauge/gravity duality', {\em Studies in History and Philosophy of Modern Physics}, 59 (2017), pp.~109-125.\\ doi:~10.1016/j.shpsb.2015.08.004. \\
\\
De Haro, S. (2016a). `Spacetime and Physical Equivalence'. To appear in {\it Space and Time after Quantum Gravity}, N. Huggett and C. Wüthrich (Eds.).\\
\\
De Haro, S., Teh, N. and Butterfield J. (2016), `Comparing dualities and gauge symmetries', {\em Studies in History and Philosophy of Modern Physics}, 59 (2017), pp.~68-80. https://doi.org/10.1016/j.shpsb.2016.03.001 \\
\\
De Haro, S., Mayerson, D., Butterfield, J.N. (2016a). `Conceptual Aspects of Gauge/Gravity Duality', {\it Foundations of Physics},  46 (11), pp.~1381-1425. doi:~10.1007/s10701-016-0037-4.\\
\\
Deser, S.~, Schwimmer, A.~(1993).  `Geometric classification of conformal anomalies in arbitrary dimensions', 
  {\it Physics Letters B}, 309, pp.~279-284
  [hep-th/9302047].\\
\\
Dieks, D., Dongen, J. van, Haro, S. de~(2015). `Emergence in Holographic Scenarios for Gravity', 
{\it Studies in History and Philosophy of Modern Physics,} 52(B), pp.~203-216. arXiv:1501.04278 [hep-th].\\
\\
Fefferman, C.~and Graham, C.R. (1985). `Conformal invariants',  in {\it Elie Cartan et les Math\'ematiques d'aujourd'hui}, Ast\'erisque, 95.\\
\\
Fefferman, C.~and Graham, C.R.~(2012). `The Ambient Metric', Annals of Mathematics Studies, number 178. Princeton University Press: Princeton and Oxford; (and at: http://arxiv.org/abs/0710.0919). \\
\\
Geroch, R.~(1972). `Structure of the Gravitational Field at Spatial Infinity', {\it Journal of Mathematical Physics}, 13, pp.~956-968; doi: 10.1063/1.1666094\\
\\
Grumiller, D.~and Riegler, M.~(2016). `Most general AdS$_{3}$ boundary conditions'. {\it Journal of High-Energy Physics}, 1610, 023. 
  doi:10.1007/JHEP10(2016)023
  [arXiv:1608.01308 [hep-th]].\\
\\
Gubser, S.~S., Klebanov, I.~R., Polyakov, A.~M.~(1998). `Gauge theory correlators from noncritical string theory',  {\it Physics Letters B}, 428, pp.~105-114 
  [hep-th/9802109].\\
\\
Hawking, S.~W.~, Perry, M.~J., and Strominger, A.~(2016). `Soft Hair on Black Holes', {\it Physical Review Letters}, 116, no.23,  p.~231301,
  doi:10.1103/PhysRevLett.116.231301
  [arXiv:1601.00921 [hep-th]].\\
\\
Henningson, M.~, Skenderis, K.~(1998). `The Holographic Weyl anomaly',  {\it Journal of High Energy Physics} 9807, 023  [hep-th/9806087].\\
\\
Horowitz, G.~and Polchinski, J.~(2006), `Gauge/gravity duality', in {\em Towards quantum gravity?}, ed. Daniele Oriti, Cambridge University Press; arXiv:gr-qc/0602037\\
\\
Imbimbo, C., Schwimmer, A., Theisen, S., Yankielowicz, S. (2000). `Diffeomorphisms and holographic anomalies', {\it Classical and Quantum Gravity}, 17(5), p.~1129.   doi:10.1088/0264-9381/17/5/322
  [hep-th/9910267].\\
\\
Ishibashi, A.~and R.~M.~Wald~(2004). `Dynamics in nonglobally hyperbolic static space-times. 3. Anti-de Sitter space-time', {\it Classical and Quantum Gravity}, 21, p.~2981;
  doi:10.1088/0264-9381/21/12/012
  [hep-th/0402184].\\
\\
 Janiszewski, S.~and Karch, A.~(2013). `Non-relativistic holography from Horava gravity',
  {\it Journal of High Energy Physics} 1302, 123.
  doi:10.1007/JHEP02(2013)123
  [arXiv:1211.0005 [hep-th]].\\
\\
Maldacena, J.~M.~(2003). `Non-Gaussian features of primordial fluctuations in single field inflationary models', {\it Journal of High-Energy Physics}, 0305, 013;
  doi:10.1088/1126-6708/2003/05/013
  [astro-ph/0210603].\\
\\
 Martelli, M., Passias, A., Sparks, J.~(2013). `The supersymmetric NUTs and bolts of holography',
{\it  Nuclear Physics} B 876, p.~810. \\doi:10.1016/j.nuclphysb.2013.04.026  [arXiv:1212.4618 [hep-th]].\\
\\
Newman, E.~T.~and Penrose, R.~(1966). `Note on the Bondi-Metzner-Sachs Group', {\it Journal of Mathematical Physics}, 7, pp.~863-870; doi: 10.1063/1.1931221\\
\\
Penrose, R.~(1963). `Asymptotic Properties of Fields and Space-Times', {\it Physical Review Letters}, 10 (2), pp.~66-68.\\
\\
Penrose, R.~(1964). `Conformal treatment of infinity', In: {\it Relativity, groups and topology}, pp.~565-584. DeWitt, B.~and DeWitt, C.~(Eds). New York and London: Gordon and Breach. Republished in: {\it General Relativity and Gravitation} (2011) 43, pp.~901-922.\\
\\
Sachs, R.~(1961). `Gravitational Waves in General Relativity. VI. The Outgoing Radiation Condition', {\it Proceedings of the Royal Society of London. Series A, Mathematical and Physical Sciences}, 264 (1318), pp.~309-338.\\
\\
Sachs, R.~(1962). `Asymptotic Symmetries in Gravitational Theory', {\it Physical Review}, 128 (6), pp.~2851-2864.\\
\\
Skenderis, K.~(2001). `Asymptotically Anti-de Sitter space-times and their stress energy tensor', {\it International Journal of Modern Physics} A, 16, p.~740.
  doi:10.1142/ S0217751X0100386X
  [hep-th/0010138].\\
\\
Skenderis, K.~(2002). `Lecture notes on holographic renormalization', {\em Classical and Quantum Gravity}, 19, pp.~5849-5876.\\
\\
Strominger, A.~(2001). `The dS / CFT correspondence', {\it Journal of High-Energy Physics}, 0110, 034;
  doi:10.1088/1126-6708/2001/10/034
  [hep-th/0106113].\\
\\
van Rees, B.~(2010). {\it Dynamics and the Gauge/Gravity Duality}, PhD thesis, University of Amsterdam. http://dare.uva.nl/record/1/326022.\\
\\
Wald, R.~M.~(1984). {\it General Relativity}. Chicago: University of Chicago Press.

\end{document}